\pdfoutput=1

\documentclass[11pt,twoside,a4paper,cmspaper,final,collab]{cms-tdr}
\def\svnVersion{303688}\def\cmsCernNo{2013-037}\def\cmsCernDate{\today}\def\cmsMessage{Published in Physics Letters B as \href{http://dx.doi.org/10.1016/j.physletb.2015.08.061}{\doi{10.1016/j.physletb.2015.08.061.}}}

\begin{document}\cmsNoteHeader{SMP-13-010}

\hyphenation{had-ron-i-za-tion}
\hyphenation{cal-or-i-me-ter}
\hyphenation{de-vices}
\RCS$Revision: 295858 $
\RCS$HeadURL: svn+ssh://svn.cern.ch/reps/tdr2/papers/SMP-13-010/trunk/SMP-13-010.tex $
\RCS$Id: SMP-13-010.tex 295858 2015-07-08 19:45:28Z alverson $

\providecommand{\qt}{\ensuremath{q_\mathrm{T}}\xspace}
\cmsNoteHeader{AN-14-113}
\title{Angular coefficients of Z bosons produced in pp collisions at $\sqrt{s}=8$\TeV and decaying to $\PGmp\PGmm$ as a function of transverse momentum and rapidity}

\date{\today}

\abstract{
Measurements of the five most significant angular coefficients, $A_{0}$ through $A_{4}$, for Z bosons
 produced in pp collisions at $\sqrt{s}=8$\TeV and decaying to $\PGmp\PGmm$ are presented as a function of the
transverse momentum and rapidity of the Z boson.
The integrated luminosity of the dataset collected with the CMS detector at the LHC
 corresponds to 19.7\fbinv.
 These measurements provide comprehensive information about the Z boson production mechanisms,
 and are compared to the QCD predictions at leading order, next-to-leading order,
 and next-to-next-to-leading order in perturbation theory. }

\hypersetup{%
pdfauthor={CMS Collaboration},%
pdftitle={Angular coefficients of Z bosons produced in pp collisions at 8 TeV and decaying to mu+mu- as a function of transverse momentum and rapidity},%
pdfsubject={CMS},%
pdfkeywords={CMS, physics, Z boson production}}

\maketitle

 We report the first measurement of the angular coefficients of Z bosons produced in pp collisions and decaying to muon pairs.
These coefficients govern the decay of the Z boson and thereby the kinematics of the lepton.
Their values follow from the vector and axial vector (V-A) structure of boson-fermion couplings.
The general structure of the lepton angular distribution in the boson rest frame is given by
\ifthenelse{\boolean{cms@external}}{
\begin{linenomath}
\begin{multline}
\label{eq:x-section}
\frac {\rd^2 \sigma } {\rd\cos\theta^{*}\rd\phi^{*}}
\propto \Bigl[(1+\cos^2\theta^{*}) +A_0 \frac{1}{2}(1-3\cos^2\theta^{*}) \\+ A_1\sin(2\theta^{*})\cos\phi^{*} + A_2\frac{1}{2}\sin^2\theta^{*}\cos(2\phi^{*})\\
 +A_3\sin\theta^{*}\cos\phi^{*} + A_4\cos\theta^{*} + A_5 \sin^2 \theta^* \sin(2\phi^*)\\+A_6\sin(2\theta^*)\sin{\phi^*} + A_7\sin{\theta^*}\sin{\phi^*}].
\end{multline}
\end{linenomath}
}{
\begin{linenomath}
\begin{equation}
\label{eq:x-section}
 \begin{split}
\frac {\rd^2 \sigma } {\rd\cos\theta^{*}\rd\phi^{*}}
\propto \Bigl[(1+\cos^2\theta^{*}) +A_0 \frac{1}{2}(1-3\cos^2\theta^{*}) + A_1\sin(2\theta^{*})\cos\phi^{*} + A_2\frac{1}{2}\sin^2\theta^{*}\cos(2\phi^{*})\\
 +A_3\sin\theta^{*}\cos\phi^{*} + A_4\cos\theta^{*} + A_5 \sin^2 \theta^* \sin(2\phi^*)+A_6\sin(2\theta^*)\sin{\phi^*} + A_7\sin{\theta^*}\sin{\phi^*}\Bigr].
 \end{split}
\end{equation}
\end{linenomath}
}
Here, $\theta^*$ and $\phi^*$ are the polar and azimuthal angles of the negatively charged lepton in the rest frame of the lepton pair.
In this analysis we choose the Collins--Soper (CS) frame~\cite{Collins:1977iv} to measure the angular coefficients $A_{i}$, considering
 the momentum of the beam proton closest in rapidity to the Z boson as the ``target momentum'' in ~\cite{Collins:1977iv}.
The parameters $A_{0}$, $A_{1}$, and $A_{2}$ are related to the polarization of the Z boson, whilst $A_{3}$ and $A_{4}$ are also sensitive to the V-A structure of the couplings of the muons.
 All angular coefficients vanish as the Z boson transverse momentum $\qt$ approaches zero
except for $A_{4}$, which is the electroweak parity violation term.

The only previous measurement of four of the angular coefficients
 was performed by the CDF Collaboration in  $\Pp\Pap$ interactions for $\qt$ up to 55\GeV~\cite{Aaltonen:2011nr}.
The angular coefficients in pp collisions are expected to differ from those in $\Pp\Pap$ collisions for several reasons.
For $\Pp\Pap$ collisions, the Z boson  production occurs predominantly via the  $\PQq\PAQq$ annihilation process,
 whilst the contribution of the qg Compton process is larger in pp collisions than  $\Pp\Pap$ collisions.
Using the \POWHEG estimation~\cite{Nason:2004rx,Frixione:2007vw,Alioli:2010xd,Alioli:2008gx} the fraction of qg process in pp collisions
 at the LHC is 47\%; it is 35\% near $\qt$ = 0 and increases to $\sim$80\% at $\qt > 100$\GeV.
For the $\PQq\PAQq$ process in the CS frame, $A_0=A_2=\qt^{2}/(M_{\Z}^2+\qt^{2})$~\cite{qq,Boer:2006eq,berger,bodek}, where $M_{\Z}$ is the Z boson mass.
For the qg Compton process $A_0=A_2\approx 5\qt^{2}/(M_{\Z}^2+5\qt^{2})$~\cite{qg}.
The relation $A_0=A_2$ is known as the Lam--Tung relation~\cite{Lam:1978zr}, reflecting the full transverse polarization of vector boson coupling to quarks, as well as rotational invariance~\cite{Rinvariance}.
Processes containing non-planar configurations (e.g., from higher order multi-gluon emission) smear the transverse polarization, leading to $A_{2} < A_{0}$~\cite{LT_ref}.  In contrast to what happens at the Tevatron, the average handedness of Z bosons is nonzero at the LHC, as for the W boson~\cite{Chatrchyan:2011ig,ATLAS:2012au,Bern:2011ie}.

The angular coefficients of Z bosons produced in pp collisions at $\sqrt{s}=8$\TeV and decaying to $\PGmp\PGmm$ are measured as a function of $\qt$ and rapidity $y$.
The data, taken with the CMS detector at the LHC, corresponds to an integrated luminosity of 19.7\fbinv.
The large Z boson event sample collected by the CMS experiment allows precision measurements of the angular distribution
for $\qt<200$\GeV and $\abs{y}<2.1$.
 The coefficients, measured as a function of $\qt$ and $\abs{y}$, are compared with three perturbative QCD predictions by \FEWZ at next-to-next-to-leading order (NNLO)~\cite{Gavin:2010az}, \POWHEG at next-to-leading order (NLO)~\cite{Nason:2004rx,Frixione:2007vw,Alioli:2010xd,Alioli:2008gx}, and \MADGRAPH at leading order (LO)~\cite{Alwall:2014hca}.

The central feature of the CMS apparatus is a
superconducting solenoid of 6\unit{m} internal diameter, providing a
magnetic field of 3.8\unit{T}. Within the solenoid
volume are a silicon pixel and strip tracker, a lead tungstate crystal
electromagnetic calorimeter, and a brass and plastic scintillator hadron
calorimeter, each composed of a barrel and two endcap
sections. Muons are measured in gas-ionization detectors embedded in the
steel flux-return yoke outside the solenoid. Extensive forward
calorimetry complements the coverage provided by the barrel and endcap
detectors. Muons are measured in the pseudorapidity range $\abs{\eta}<
2.4$, with detection planes made using three technologies: drift tubes,
cathode strip chambers, and resistive plate chambers.
A more detailed description of the CMS
detector, together with a definition of the coordinate system and
 the relevant kinematic variables, can be found in Ref.~\cite{Chatrchyan:2008zzk}.

 Matching muons to tracks measured in the silicon tracker results in a relative $\pt$
resolution for muons with $20 <\pt < 100$\GeV of 1.3--2.0\% in the
barrel and better than 6\% in the endcaps.
 A particle-flow (PF) event reconstruction algorithm~\cite{CMS:2009nxa,CMS:2010byl}
is used in this analysis. It consists of reconstructing and identifying
each single particle with an optimized combination of all subdetector
information.
A trigger for single isolated muon is used, requiring $\pt> 24$\GeV and $\abs{\eta} < 2.1$.
The leading in $\pt$ reconstructed muon is matched to the muon selected by the trigger.

The signal process is simulated using the \MADGRAPH~1.3.30 generator~\cite{Alwall:2014hca} with zero to four additional
jets, interfaced with \PYTHIA v6.4.24~\cite{Sjostrand:2006za} with the Z2*
tune~\cite{z2tune}. The matching between the matrix element calculation and the parton shower is performed with the \kt-MLM algorithm~\cite{Alwall:2007fs}. The CTEQ6L1~\cite{Pumplin:2002vw} parton distribution functions (PDF) are used for the event generation. Multiple-parton interactions are simulated by \PYTHIA.
The \POWHEG generator~\cite{Nason:2004rx,Frixione:2007vw,Alioli:2010xd,Alioli:2008gx}
 interfaced with \PYTHIA (same version used for \MADGRAPH) and the CT10 PDF set~\cite{ct10_pdf} are used as an alternate to
test any model dependence in the shapes of the angular distributions.

Background simulations are performed with \MADGRAPH
(W+jets, $\ttbar$, $\PGt\PGt$), \POWHEG (single top quark~\cite{Alioli:2009je,Re:2010bp}), and \PYTHIA (WW, WZ, ZZ).
 The normalizations of the inclusive Drell--Yan, W boson~\cite{Gavin:2010az}, and
$\ttbar$~\cite{Czakon:2013goa} distributions are set using NNLO cross sections.
 For single top quark production a higher order (approximate NNLO~\cite{Kidonakis:2012db}) inclusive
cross section is used.  The generated events are passed through a
detector simulation based on \GEANTfour~\cite{Agostinelli:2002hh}.

Each muon candidate is required to be reconstructed in the muon
detectors and in the inner tracker, and the global track fit is
required to have a reduced $\chi^2 < 10$. The vertex with the highest sum of $\pt^2$
 for associated tracks is defined as the primary vertex.
The distance of the muon candidate trajectories
with respect to the primary vertex must be smaller than 2\mm in the
transverse plane and 5\mm along the beam axis.
The leading (subleading) muon is required to have $\pt > 25\,(10)$\GeV
and $\abs{\eta} < 2.1~(2.4)$. In order to suppress background events, the muons
are required to be isolated from nearby particles.
The relative isolation is calculated as the ratio of the scalar sum of $\pt$ of all PF candidates
from the same primary vertex, within a cone of $\Delta
R=\sqrt{\smash[b]{(\Delta\eta)^2+(\Delta \phi)^2}}<0.4$, and the $\pt$ of the muon.
 For the leading (subleading) muon in $\pt$, the relative isolation must be less than 0.12 (0.5).
 Oppositely charged muon pairs with an invariant mass in the range 81--101\GeV are selected. In the
rare case that more than two muons are selected, the muon pair with invariant mass closest to
the Z boson mass is chosen.
The muon pair must satisfy $\abs{y} < 2.1$ since at higher $\abs{y}$ the acceptance varies rapidly.
After the event selection, 4.3$\times 10^{6}$ events with Z boson candidates remain for $\abs{y} < 1.0$ and 2.5$\times 10^{6}$ events for $1.0 < \abs{y} < 2.1$.

A ``tag-and-probe'' method~\cite{Chatrchyan:2012xi} is used to measure
the efficiencies for track reconstruction, trigger, muon isolation, and
 muon identification in data and simulation.
Efficiency corrections are applied as multiplicative scale factors to the simulation values.
The efficiency for track reconstruction is measured in bins of
$\eta$ since the $\pt$ dependence is weak. The trigger efficiency is determined in bins of
$\pt$ and $\eta$, separately for $\PGmp$ and $\PGmm$. The
identification efficiency is measured in bins of $\pt$ and $\eta$.
Since the subleading muon can point in the direction of the hadronic activity,
a looser isolation requirement is used and its efficiency is measured
as a function of $\qt$, $\cos\theta^*$, and $\phi^*$.
The efficiency of the isolation requirement for the leading muon is measured as a function of $\pt$ and $\eta$ of the muon,
 as detector effects relate to these variables more directly than to the Z boson $\qt$ and $y$.

After  event selection, the background contribution ranges from $\sim$0.1\% at low $\qt$ to $\sim$1.5\% at high $\qt$.
 The yields of the backgrounds from
$\ttbar$, $\PGt\PGt$, WW, tW,
 and W+jets production are estimated from data using lepton flavor universality.
 Most of these backgrounds typically have two prompt leptons, which may have the same
flavor. The W+jets background is flavor asymmetric, but its contribution is small.
We assume that the ratio of the number of
oppositely charged background  $\PGm\PGm$ and $\Pe\PGm$ events is the same in data and
simulation. We use the ratio of the $\Pe\PGm$ yields in data and
simulation after applying muon and electron selection criteria~\cite{Chatrchyan:2012xi,CMS:2013hoa} to normalize the simulation to data.

The acceptance and the efficiency at the event level vary in
$\cos\theta^*$ and $\phi^*$, and strongly with $\qt$ and
$y$. In order to avoid a bias in the acceptance due to the modeling of the Z boson kinematics,
 the simulation is reweighted in fine bins of $\qt$ and
$y$ to match the background-subtracted data distribution.
The weights are determined at the reconstruction level and applied at the generator level.
The weighting is iterated four times, with negligible change between the second and fourth iteration.

The angular coefficients are measured in eight bins of $\qt$ and two bins of $\abs{y}$,
 by fitting the two-dimensional ($\cos\theta^*$, $\phi^*$) distribution in data with a linear combination of templates.
These templates are built for each coefficient $A_i$ by reweighting the simulation at generator
level to the corresponding angular distribution, as given in
Eq.~\eqref{eq:x-section}.
The templates are based on reconstructed muons, and thereby incorporate the effects of resolution,
 efficiency and acceptance.
 A template is also  built for the term $(1+\cos^2{\theta^*})$ of Eq.~\eqref{eq:x-section}.
 An additional template, with shape and normalization fixed,
is developed for fitting the backgrounds. A binned maximum-likelihood method with Poisson uncertainties is employed for the fit.
  The angular coefficients $A_5$, $A_6$, and $A_7$ are predicted to be very small; they are set to zero and excluded from the fit.
 Since $A_{0}$ through $A_{4}$ are sign
invariant in $\phi^{*}$, the absolute value $\abs{\phi^{*}}$ is used.
The fit is made in 12$\times$12 equidistant bins in $\cos\theta^{*}$ and $\abs{\phi^{*}}$.
The statistical uncertainties from the fit are confirmed by comparison with pseudo-experiments.

To test the robustness of the result with respect to the analysis method and trigger effect,
 the angular coefficients $A_0$, $A_2$, $A_3$, and $A_4$ are also measured by
an independent analysis similar to that reported in Ref.~\cite{Aaltonen:2011nr}, where
one-dimensional (1D) templates produced using  \POWHEG are fitted to the distributions in
$\cos\theta^*$ and $\abs{\phi^{*}}$.
The 1D fit analysis is performed iteratively, so as to be unbiased with respect to the assumed templates and to possible
 correlations between $\cos\theta^{*}$ and $\abs{\phi^{*}}$.
The analysis differs in the triggers,  estimation of backgrounds, simulation, and selection criteria.
The 1D fit analysis uses a sample that requires a dimuon trigger with asymmetric muon $\pt$ thresholds of 17 and 8\GeV.
Both results are consistent within their total systematic uncertainties, excluding uncertainties common to both analyses.

\begin{figure*}[!ht]
 \centering
\includegraphics[width=0.45\textwidth] {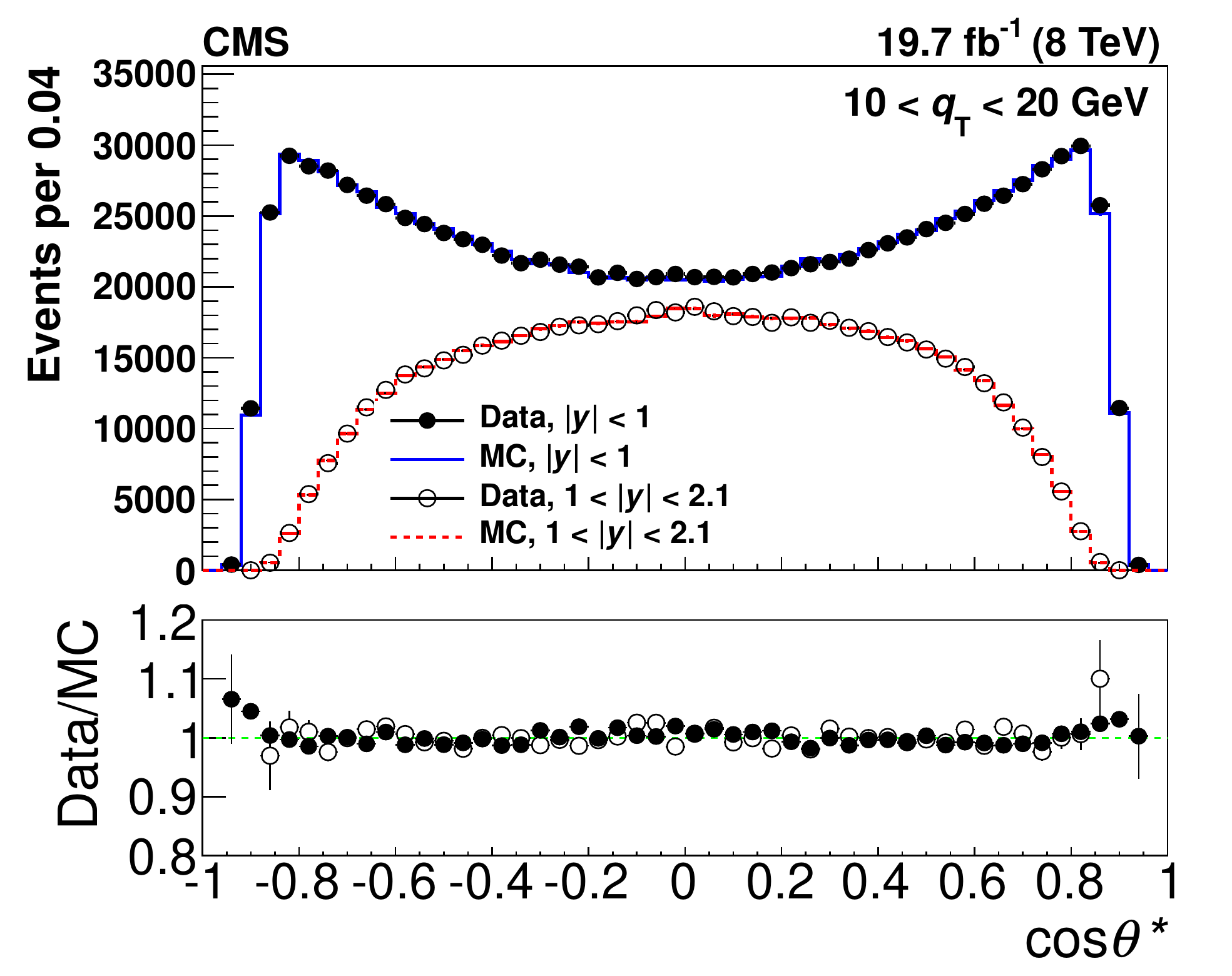}
\includegraphics[width=0.45\textwidth] {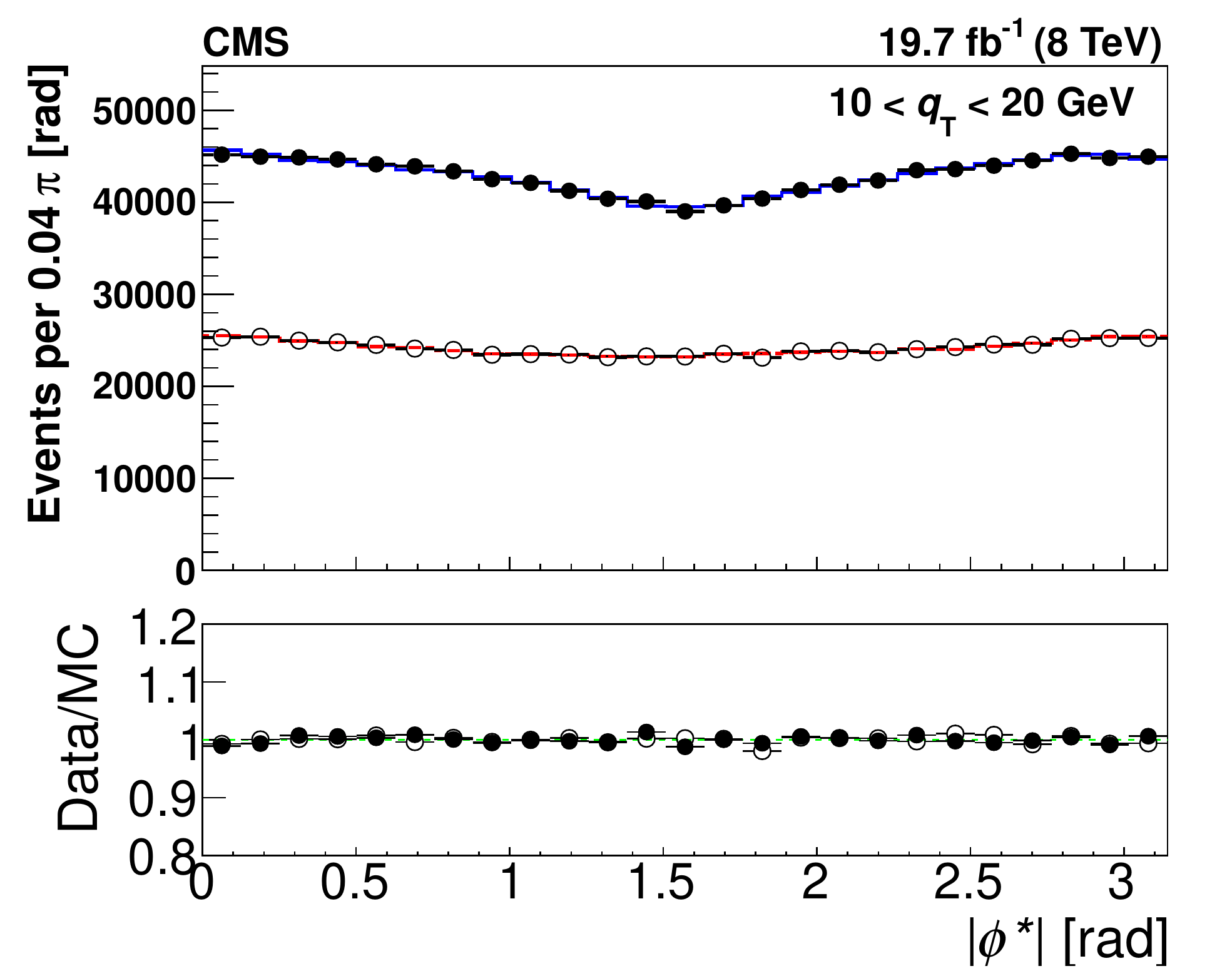}
\includegraphics[width=0.45\textwidth] {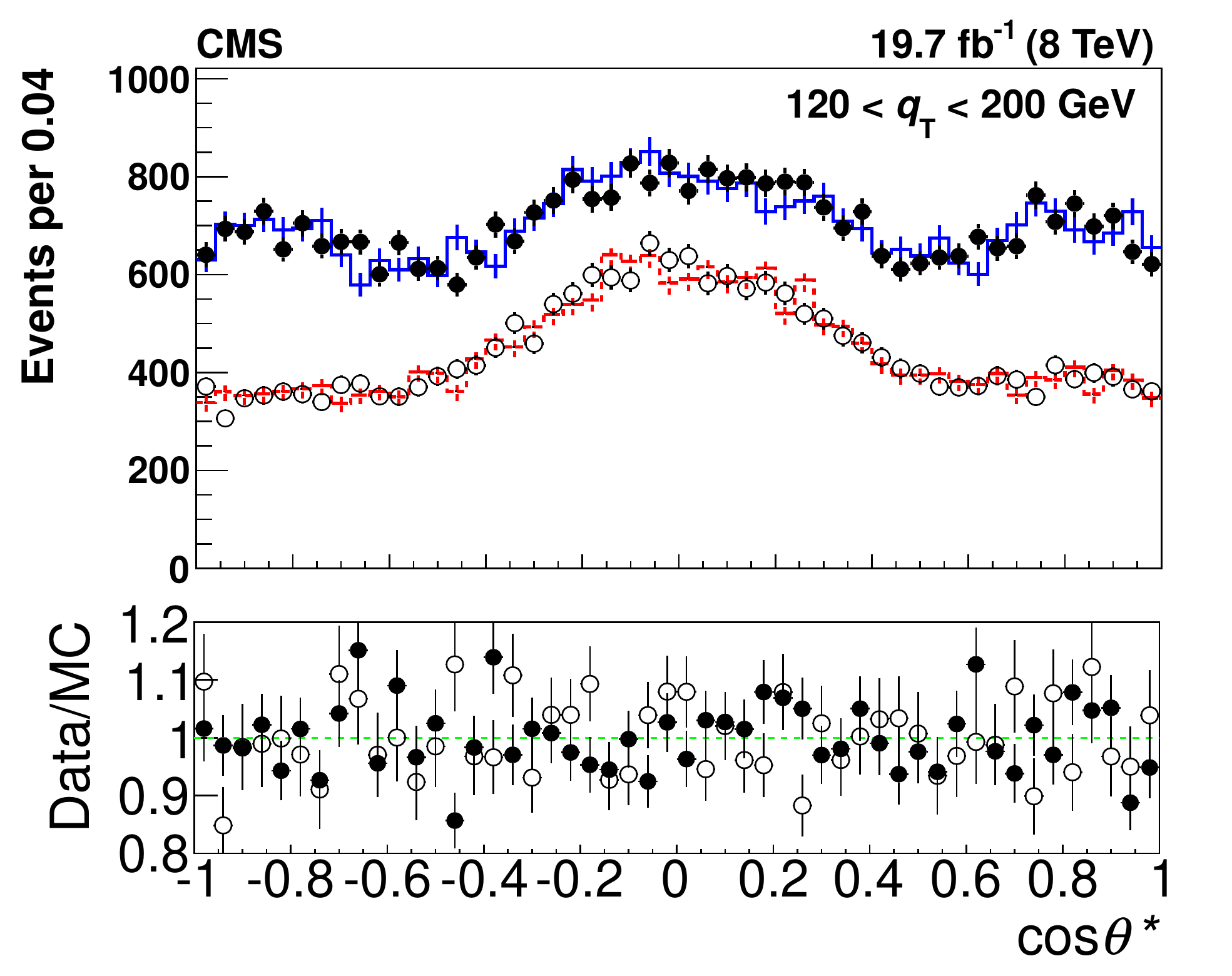}
\includegraphics[width=0.45\textwidth] {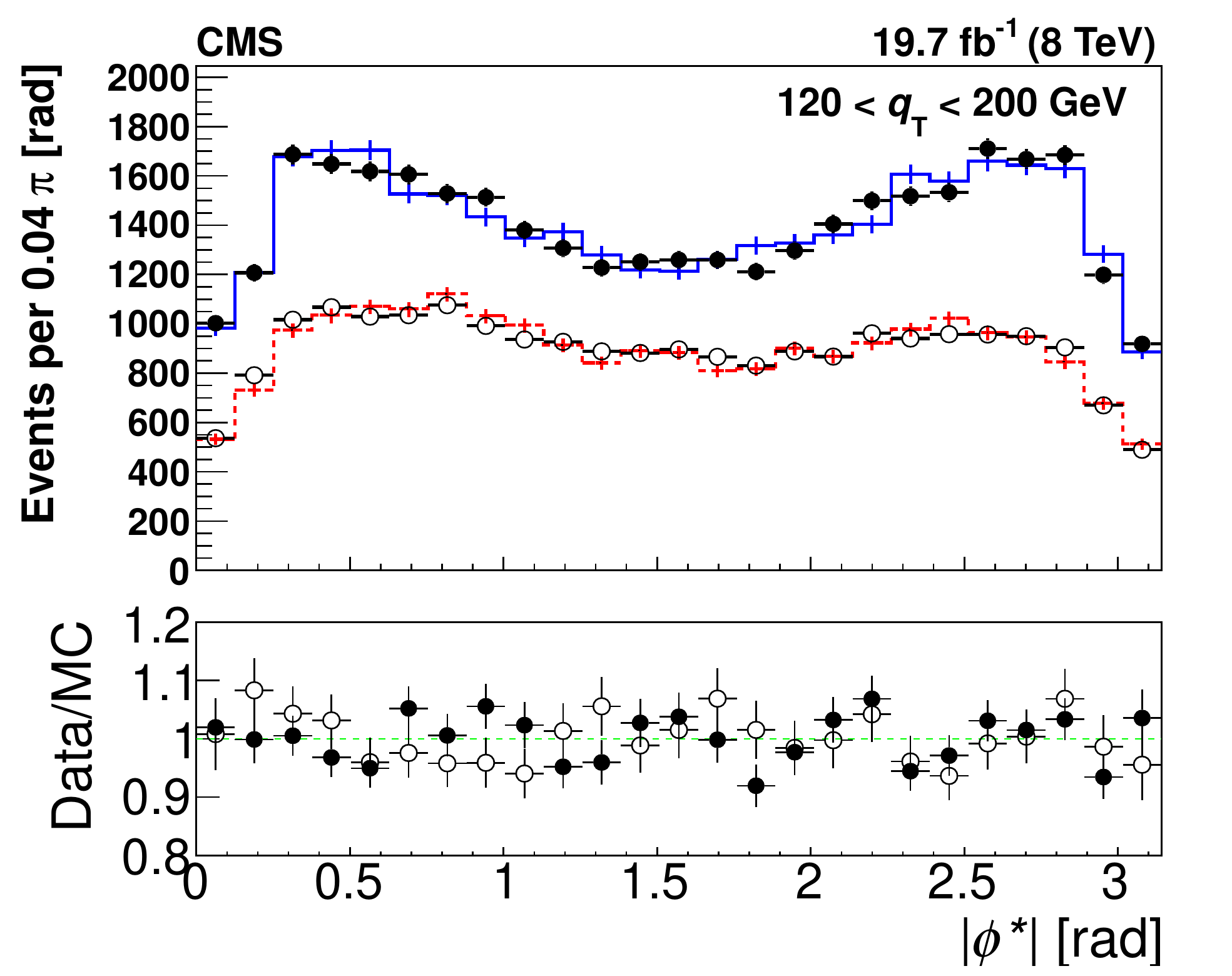}

\caption{A few examples of the observed 1D angular distributions in  $\cos\theta^{*}$ (left) and $\abs{\phi^{*}}$ (right) compared to the MC simulation using the best fit values of the angular coefficients. The top (bottom) plots show the distributions for $10<\qt<20$\GeV ($120<\qt<200$\GeV), a region where $A_0$ and $A_2$ are small (large).
The background-subtracted data points are shown with filled (open) circles for $\abs{y}<1$ ($1<\abs{y}<2.1$), whilst the corresponding MC results are shown with the solid (dashed) lines. Vertical bars represent the statistical uncertainties. The lower panels show the data-to-MC ratios.
}
\label{fig:angular_dist}
\end{figure*}

Some examples of the measured $\cos\theta^{*}$ and $\abs{\phi^{*}}$ distributions from the 1D analysis are given in Fig.~\ref{fig:angular_dist}. The measured and simulated distributions are shown together using the best fit values of the angular coefficients.
The shape of the $\cos\theta^{*}$ distribution changes with $\qt$ and $\abs{y}$ because the acceptance and efficiency in $\cos\theta^{*}$ depend strongly on these two variables. For $\abs{\phi^{*}}$, the shape of the distribution changes moderately with $\qt$, and is almost insensitive to $\abs{y}$.
The comparison of data and simulation shown in Fig.~\ref{fig:angular_dist} gives confidence that the acceptance and efficiency dependences are correctly modeled in the simulation.

 Several sources of systematic uncertainties are taken into account. The most significant source is
 the muon efficiency that includes the trigger, track reconstruction, isolation, and identification.
 The statistical uncertainties of the measured efficiency scale
factors are taken into account by simulating 500 pseudo-experiments in which the templates are
reformed, each time varying the scale factors randomly within the given uncertainty.
The systematic uncertainties in the extraction of the efficiency (e.g., background estimates) are also included.
Another significant uncertainty stems from the statistical
precision of the templates, which is estimated using pseudo-experiments.
 The pileup uncertainty is estimated by varying the
cross section of the minimum bias events by ${\pm}5\%$. The muon momentum
bias is measured in data and simulation, and corresponding
corrections are applied~\cite{rochcor}. The statistical uncertainties in the muon
momentum correction factors are propagated to a systematic uncertainty using pseudo-experiments. In
addition, a systematic uncertainty is assessed to take into account
possible global offsets from the peak position of the Z boson mass.
 The systematic uncertainties for the background
are estimated by varying the normalization scale factor of the $\Pe\PGm$
sample by 10\% and the yields of WZ and ZZ events by 50\%.
The statistical precision of the iterative reweighting is determined using pseudo-experiments.
The difference between the last two iterations is assigned as additional systematic uncertainty.
  The effect of final-state
radiation is taken into account by adding the energy of photons within a cone of radius 0.1
around the muon direction~\cite{Khachatryan:2010xn}. Weights are applied to the simulation to
reflect the difference between a soft-collinear approach and the exact
O($\alpha_\mathrm{QED}$) result and the reconstructed template is rebuilt using the weighted simulation.
The difference between templates is used to estimate the systematic uncertainty from final-state radiation.
 Finally, the acceptance uncertainty, related to the values of $A_{i}$ assumed in the simulation, is estimated by reweighting
 with the fitted values of $A_{i}$, and the difference in results is included as a systematic uncertainty.
Generally, the statistical uncertainties dominate in the highest bins in $\qt$,
whilst the systematic uncertainty in the efficiency tends to be the most important elsewhere.

\begin{figure*}[!ht]
 \centering
\includegraphics[width=0.329\textwidth] {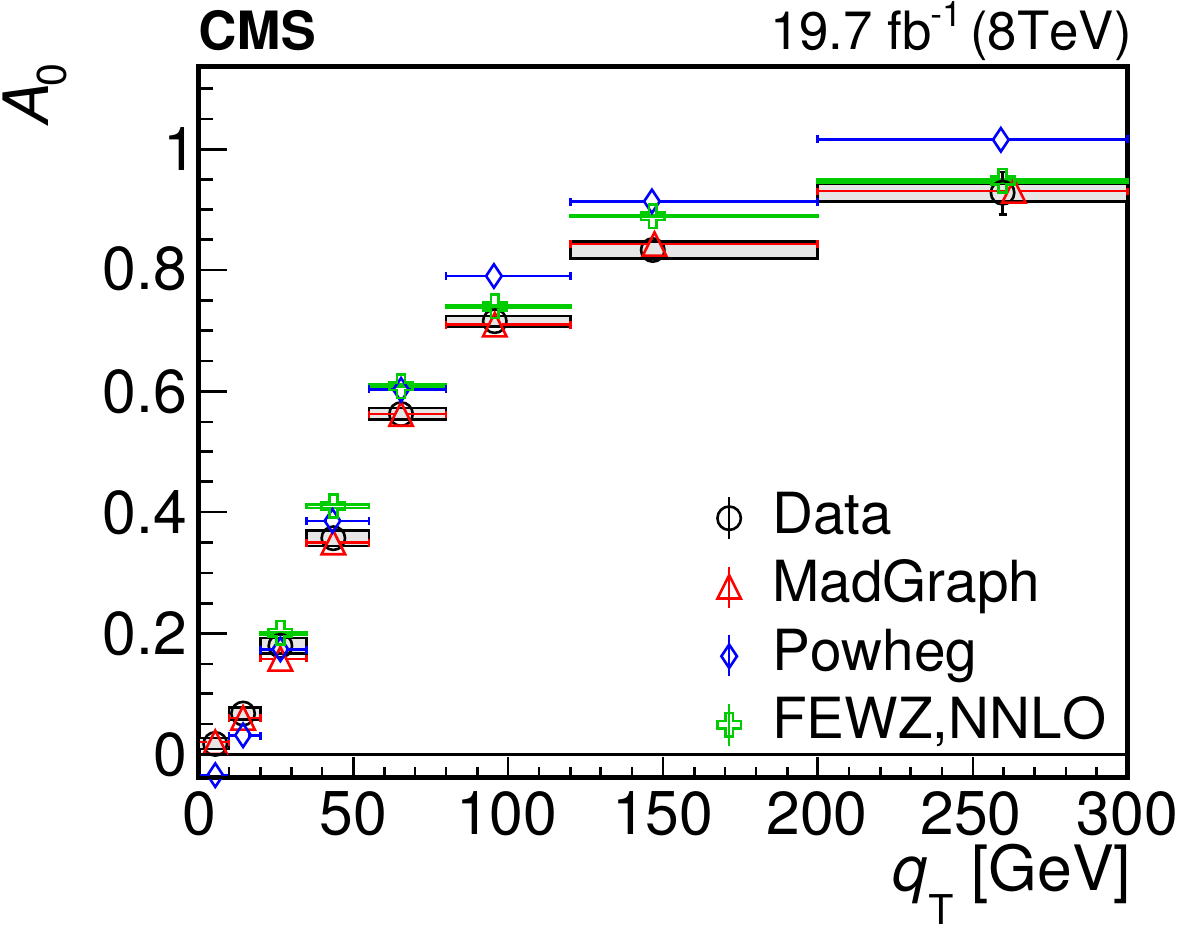}
\includegraphics[width=0.329\textwidth] {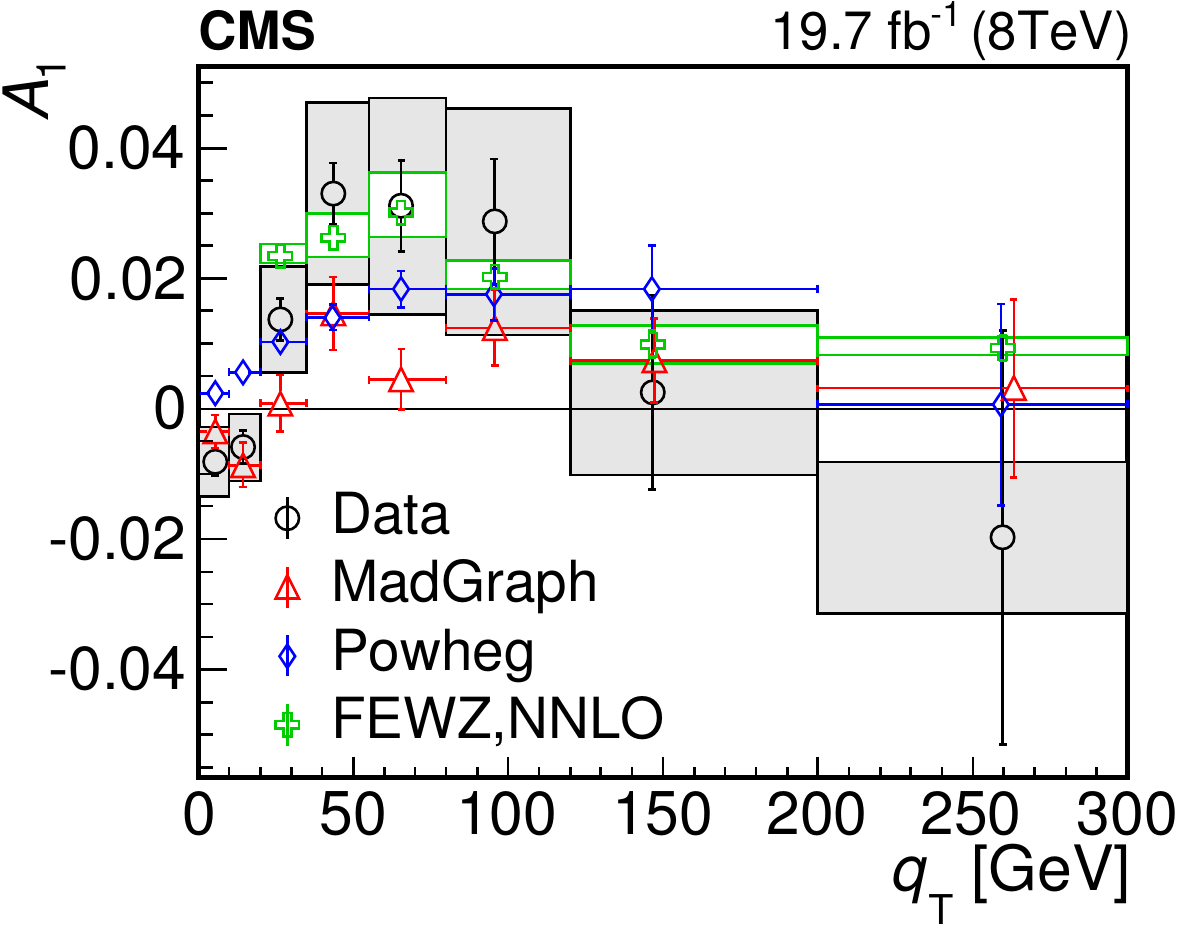}
\includegraphics[width=0.329\textwidth] {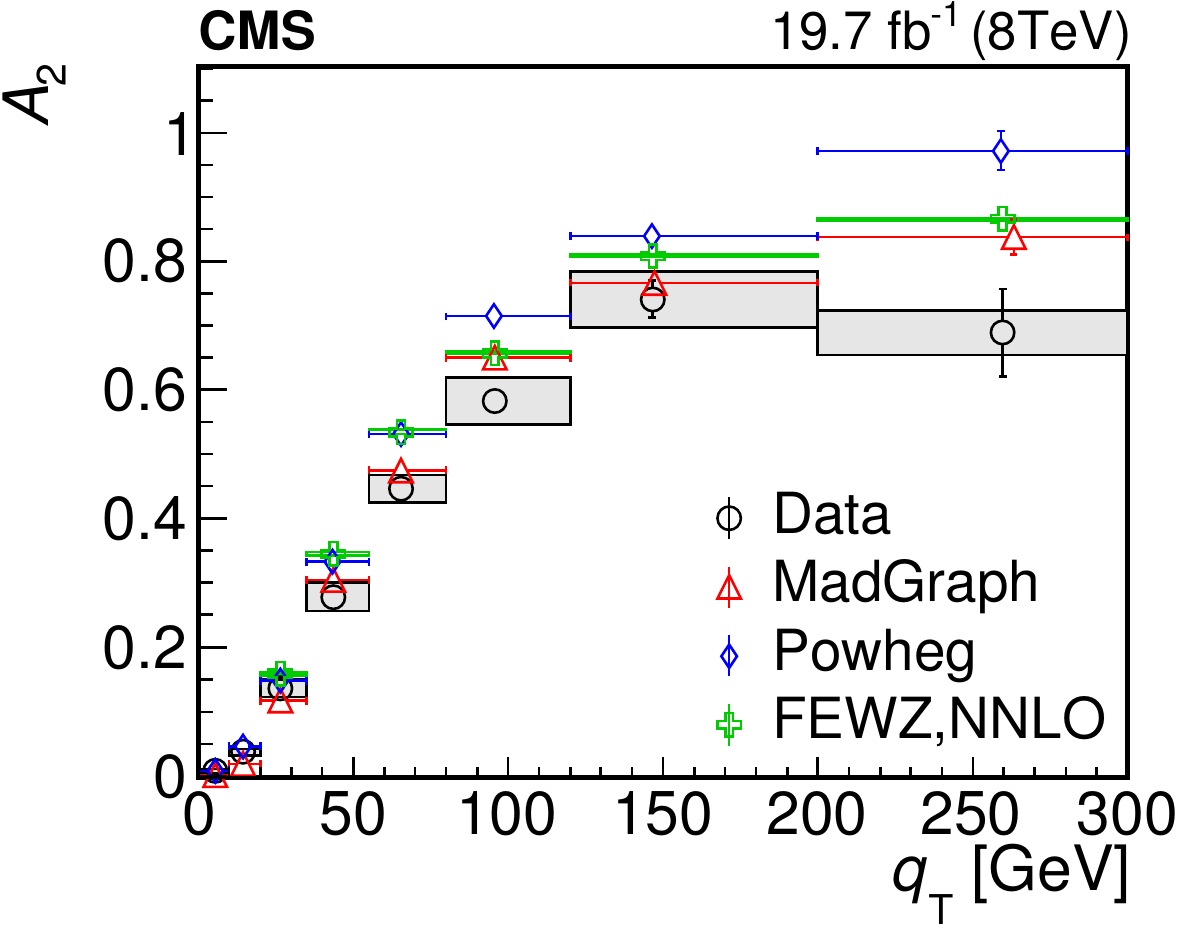}
\includegraphics[width=0.329\textwidth] {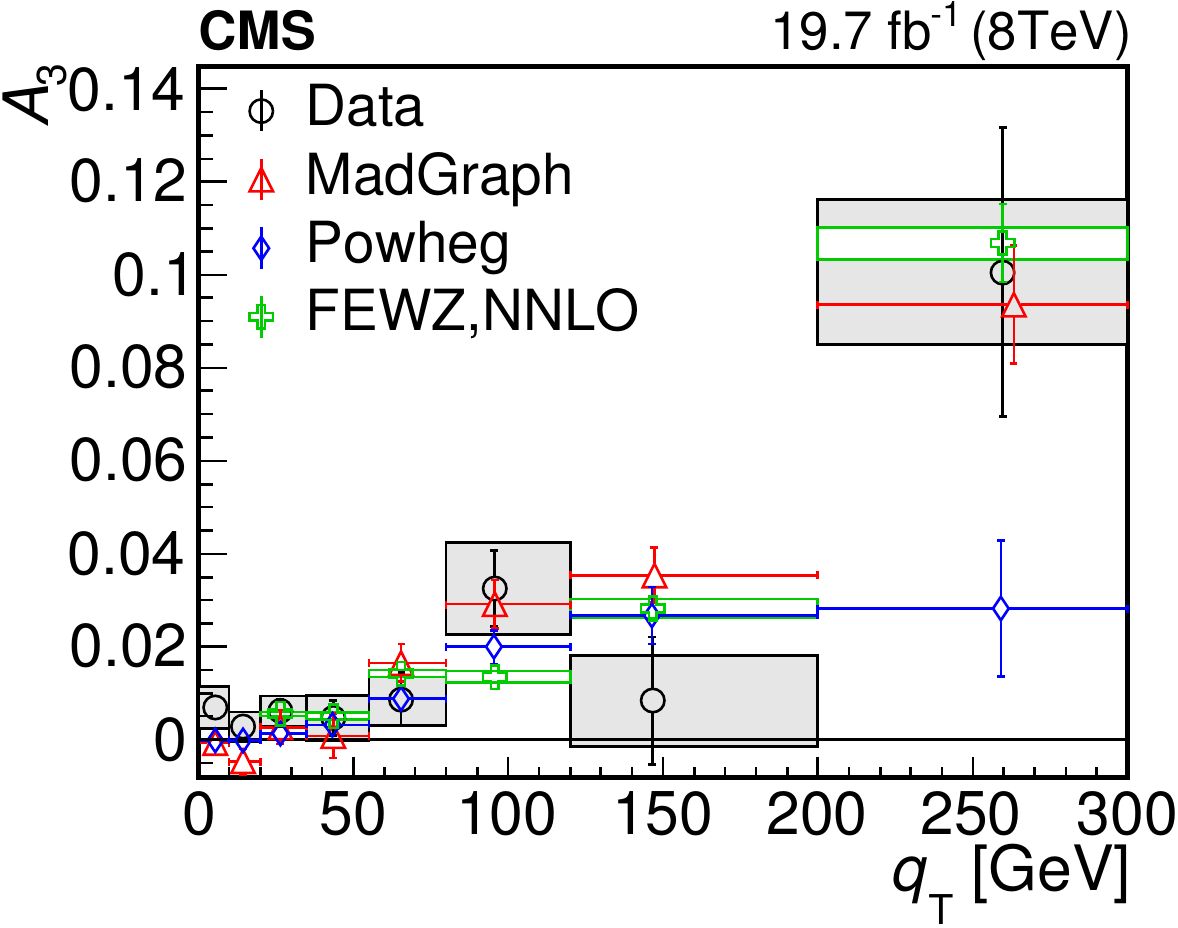}
\includegraphics[width=0.329\textwidth] {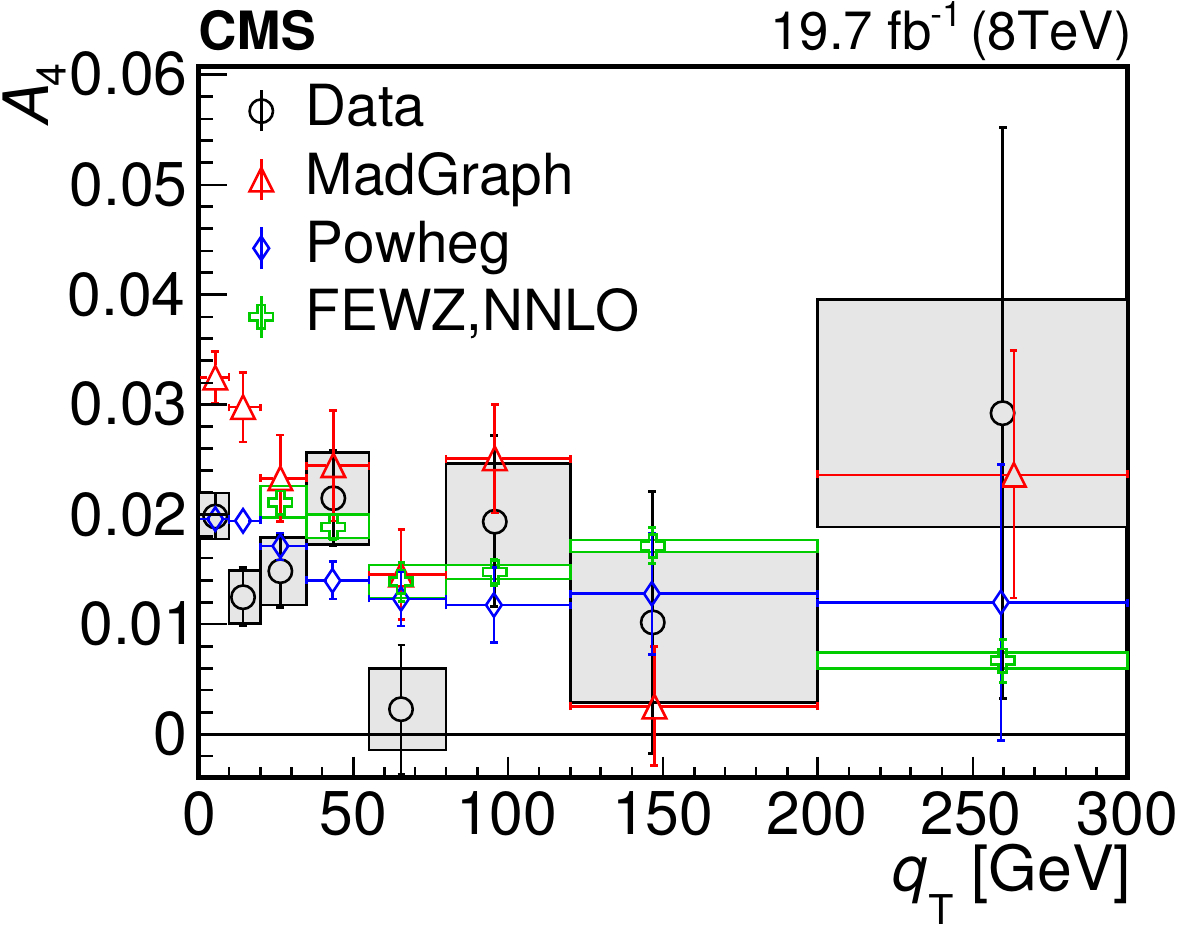}
\includegraphics[width=0.329\textwidth] {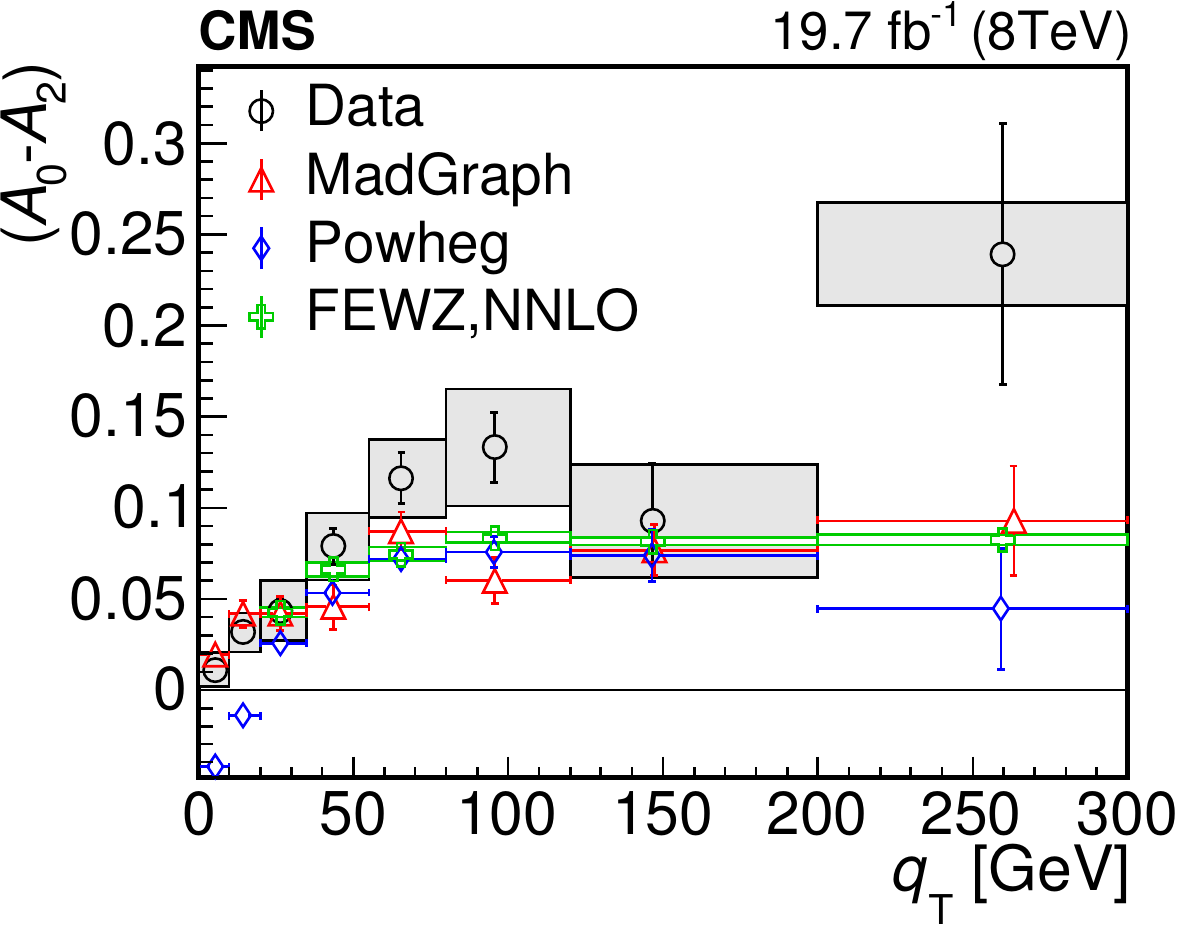}

\caption{Comparison of the five angular coefficients $A_i$ and $A_0-A_2$ measured in the Collins--Soper frame in bins of $\qt$ for $\abs{y}<1$. The circles show the measured results. The vertical bars represent the statistical uncertainties and the
boxes the systematic uncertainties of the measurement. The triangles show the predictions from \MADGRAPH, the diamonds from \POWHEG, and the crosses from \FEWZ at NNLO.  The boxes at the \FEWZ values indicate the PDF uncertainties.}
\label{fig:pol_res_cs_Y0}
\vspace{0.5cm}
 \centering
\includegraphics[width=0.329\textwidth] {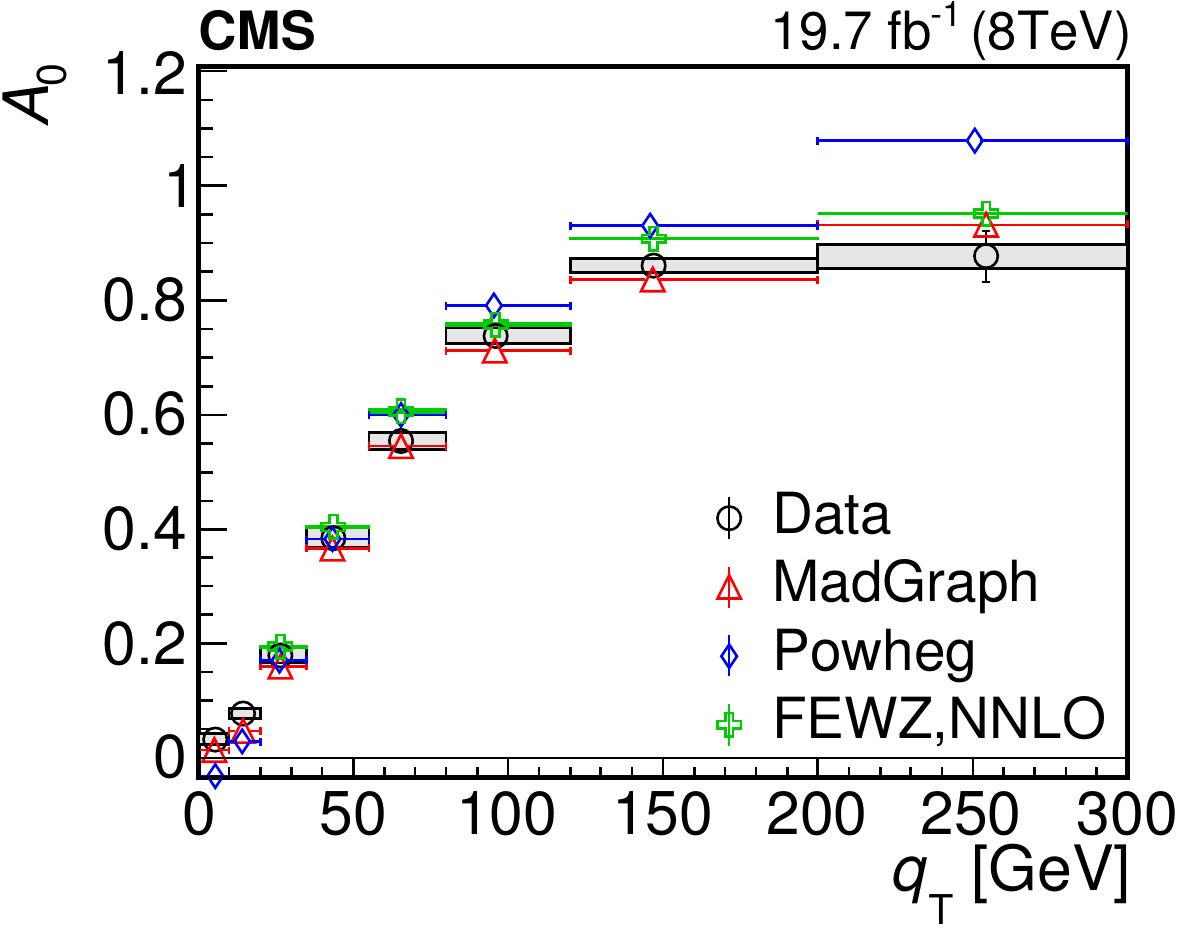}
\includegraphics[width=0.329\textwidth] {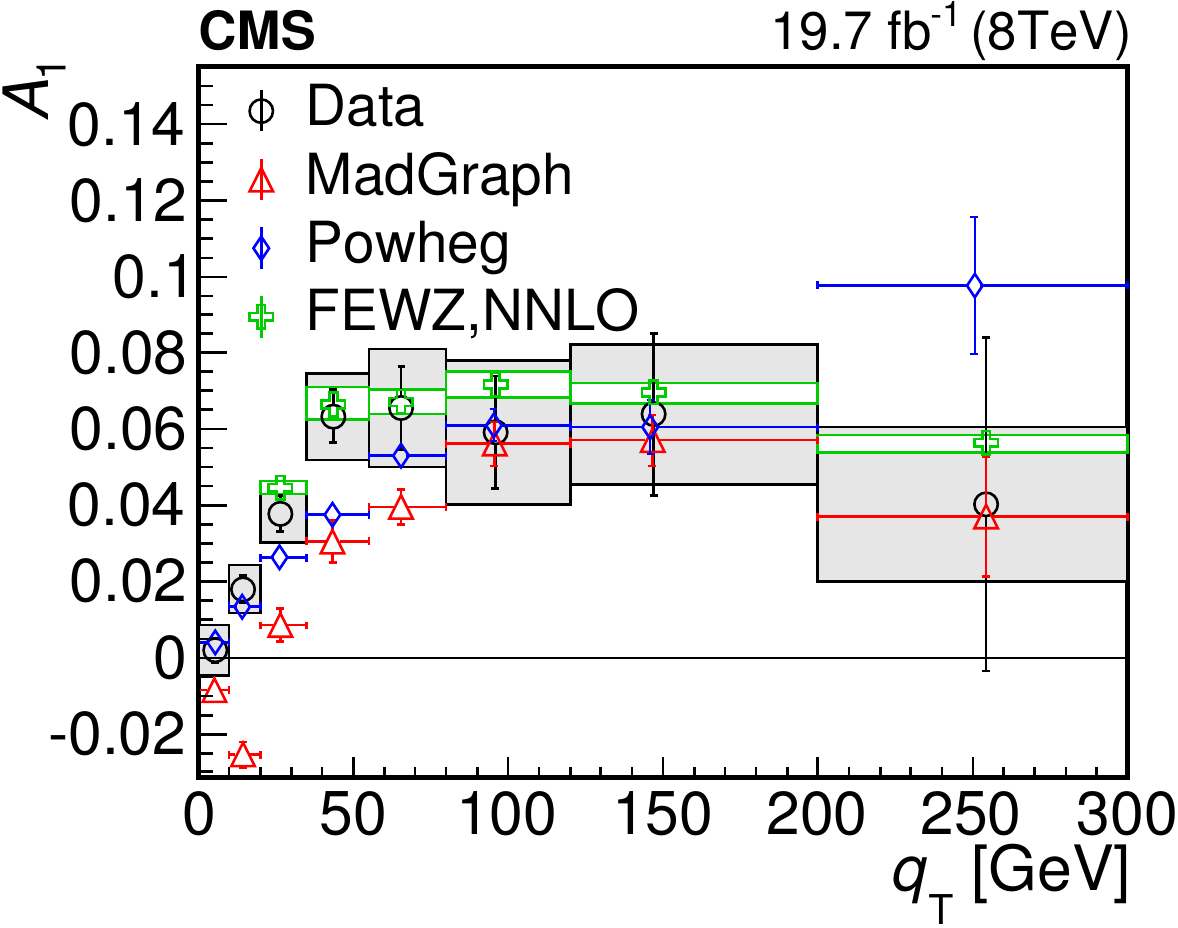}
\includegraphics[width=0.329\textwidth] {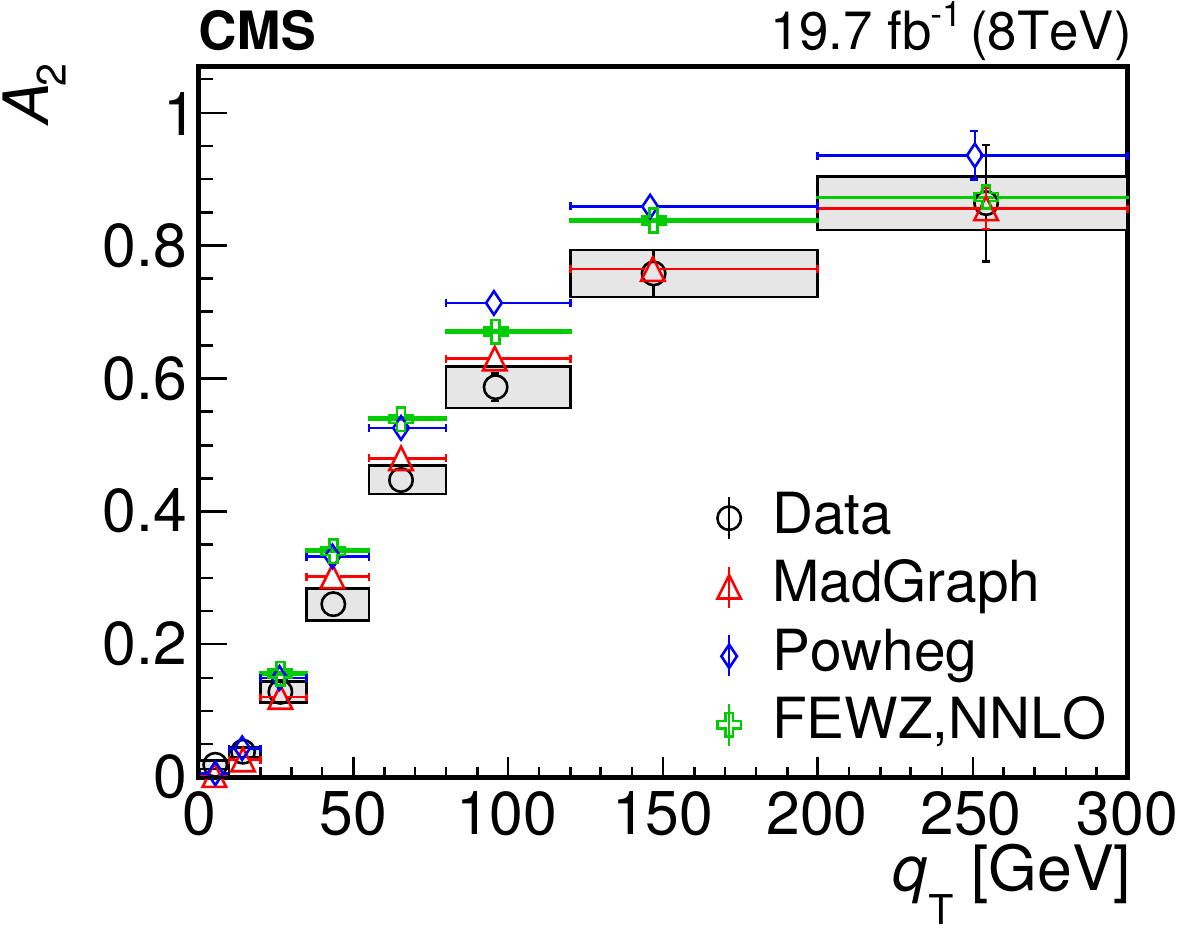}
\includegraphics[width=0.329\textwidth] {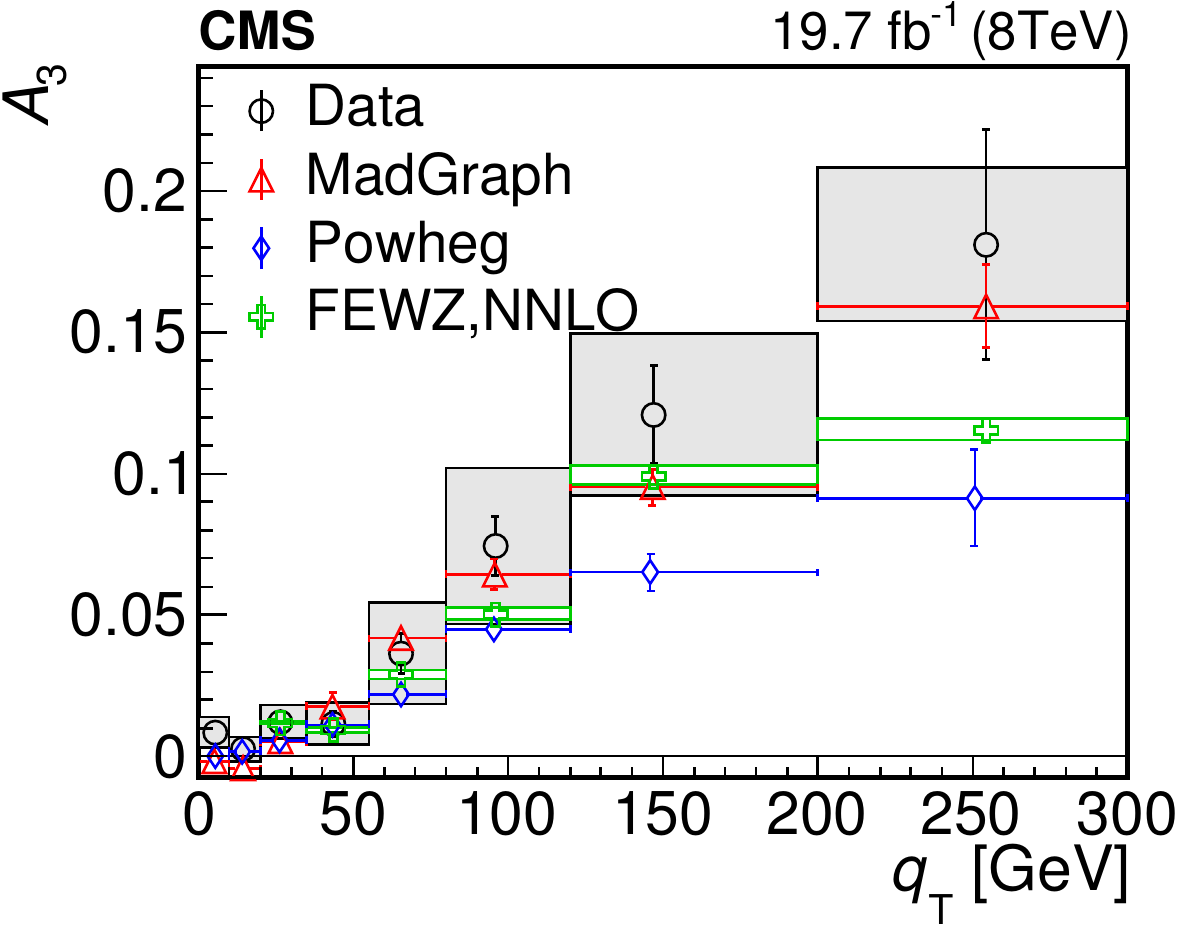}
\includegraphics[width=0.329\textwidth] {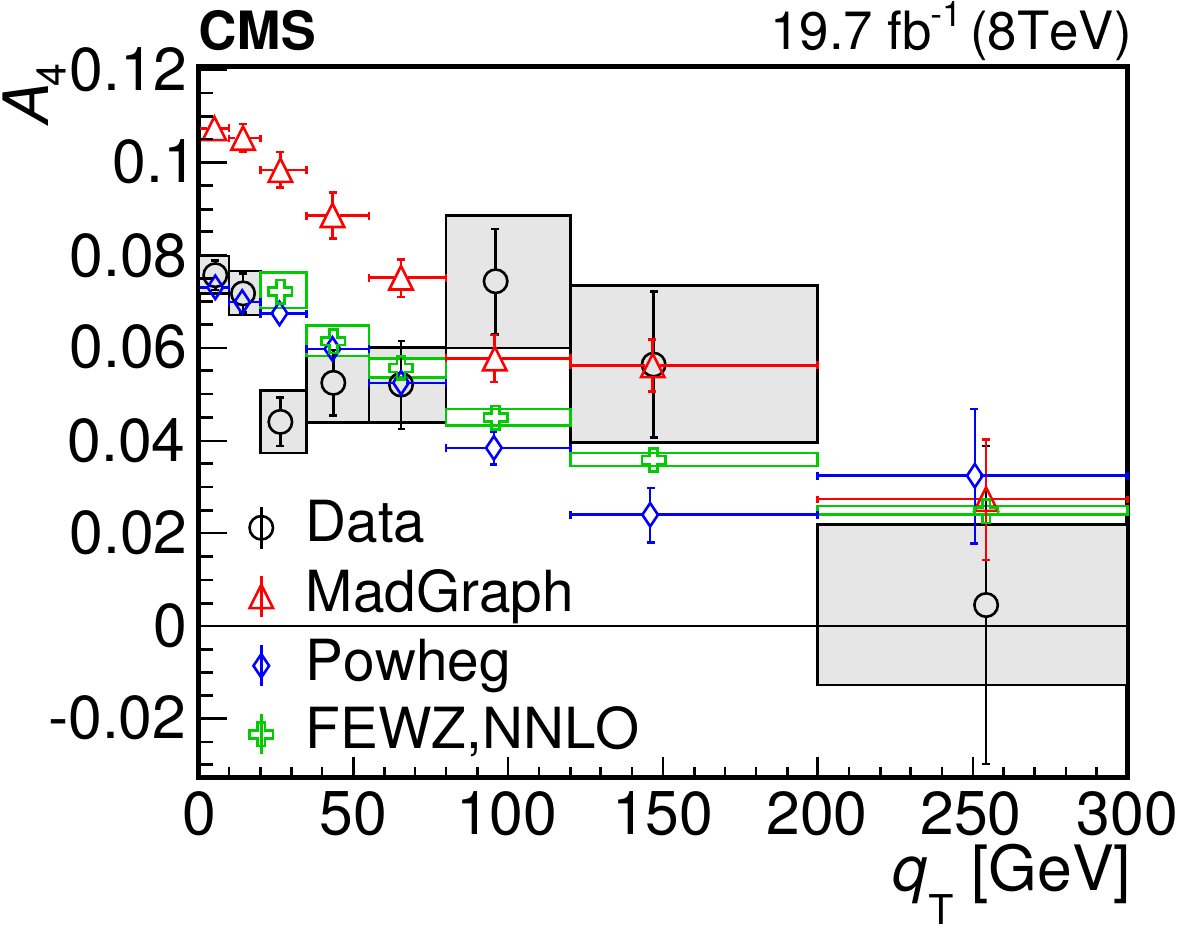}
\includegraphics[width=0.329\textwidth] {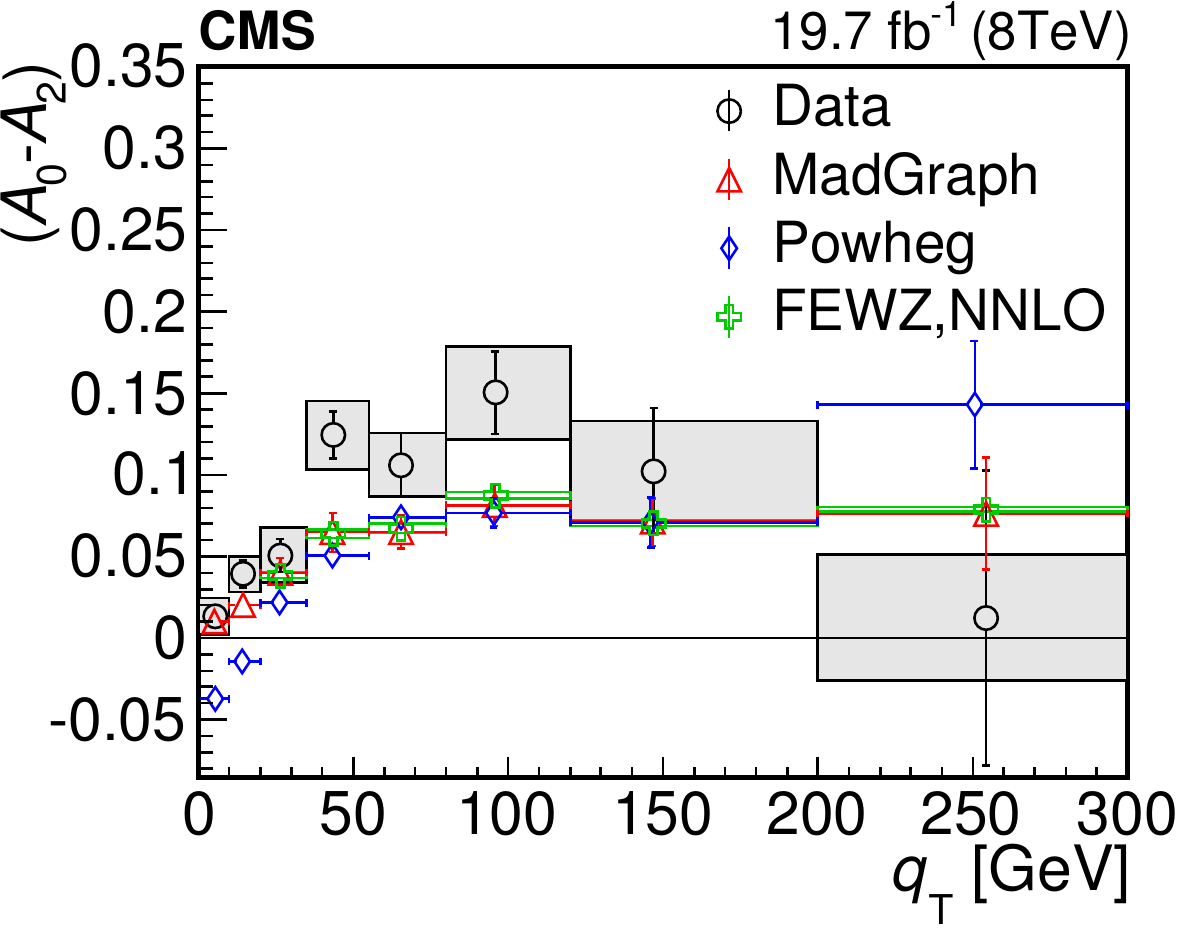}

 \caption{Comparison of the five angular coefficients and $A_0-A_2$ under the same conditions as Fig.~\ref{fig:pol_res_cs_Y0}, for the rapidity bin $1<\abs{y}<2.1$.
}
\label{fig:pol_res_cs_Y1}
\end{figure*}

\begin{figure*}[p]
 \centering
\includegraphics[width=0.36\textwidth]{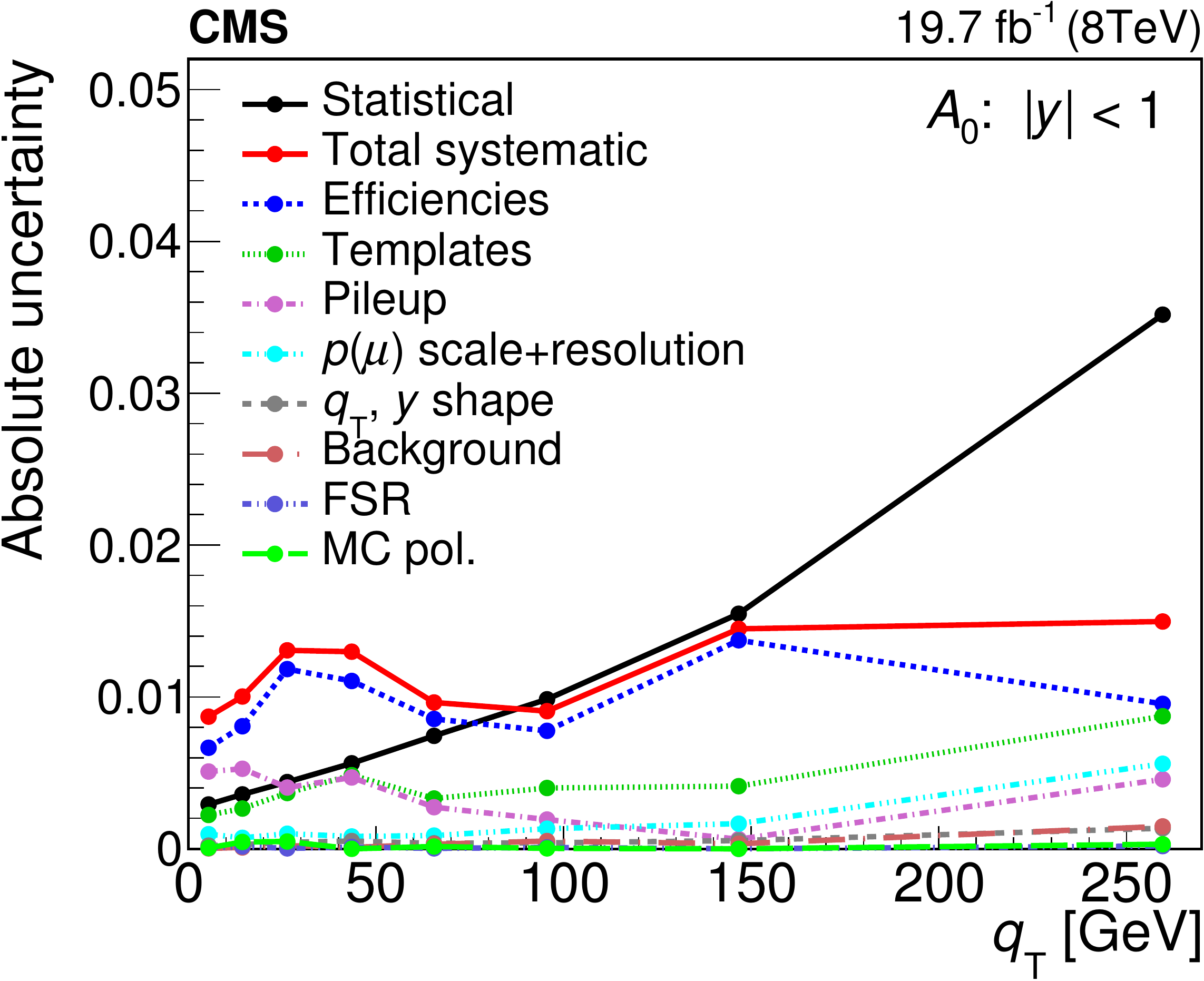}
\includegraphics[width=0.36\textwidth]{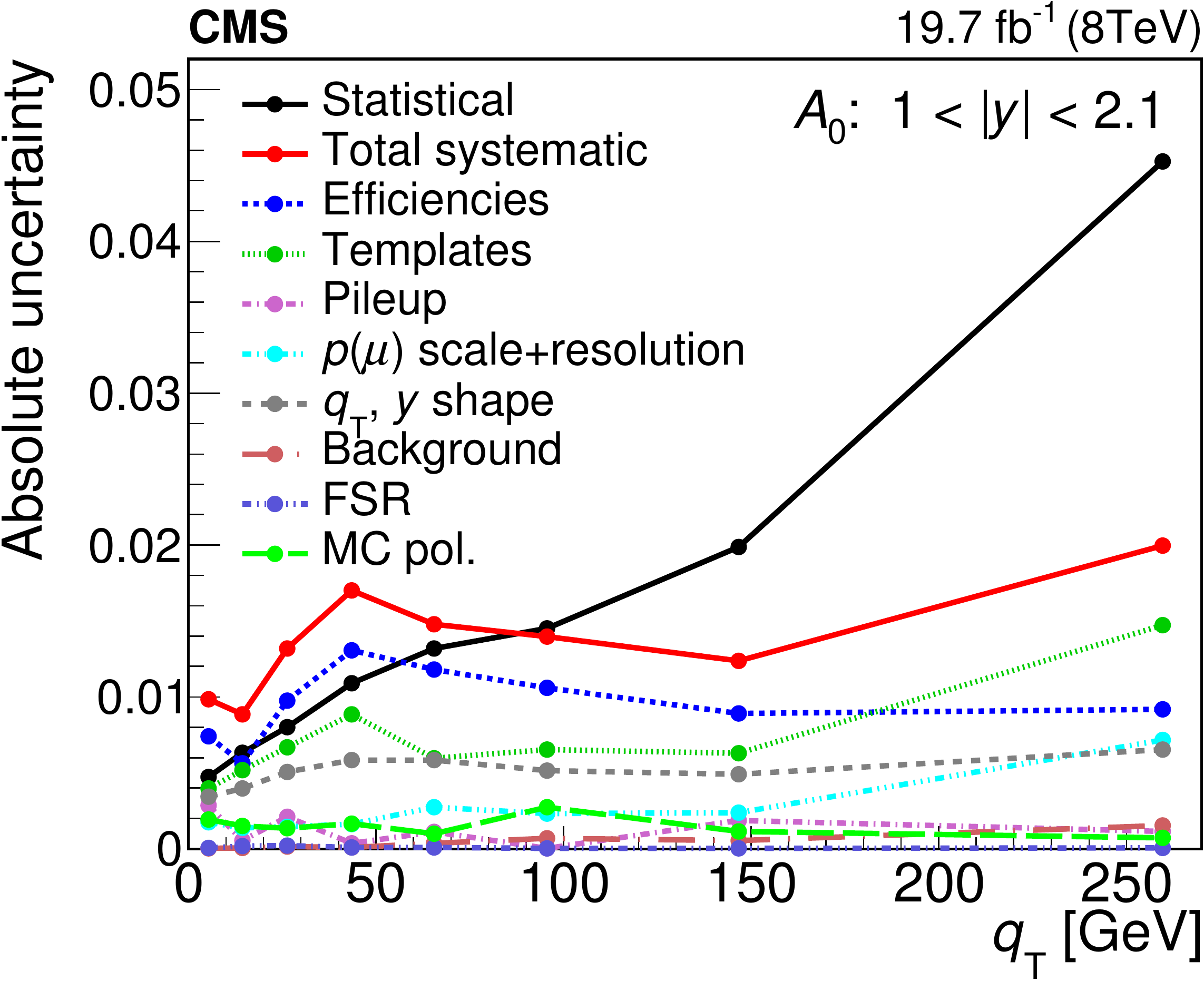}\\
\includegraphics[width=0.36\textwidth]{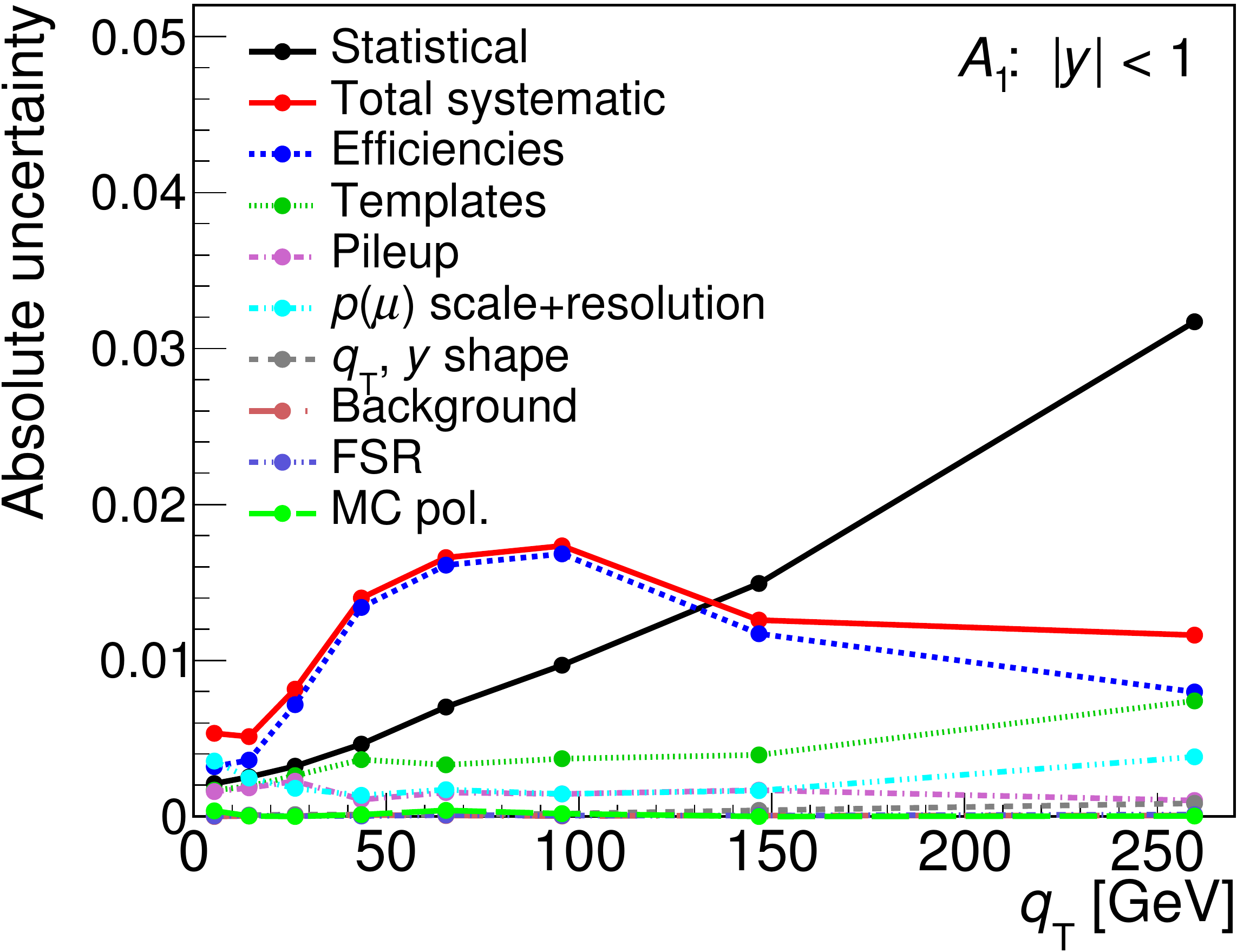}
\includegraphics[width=0.36\textwidth]{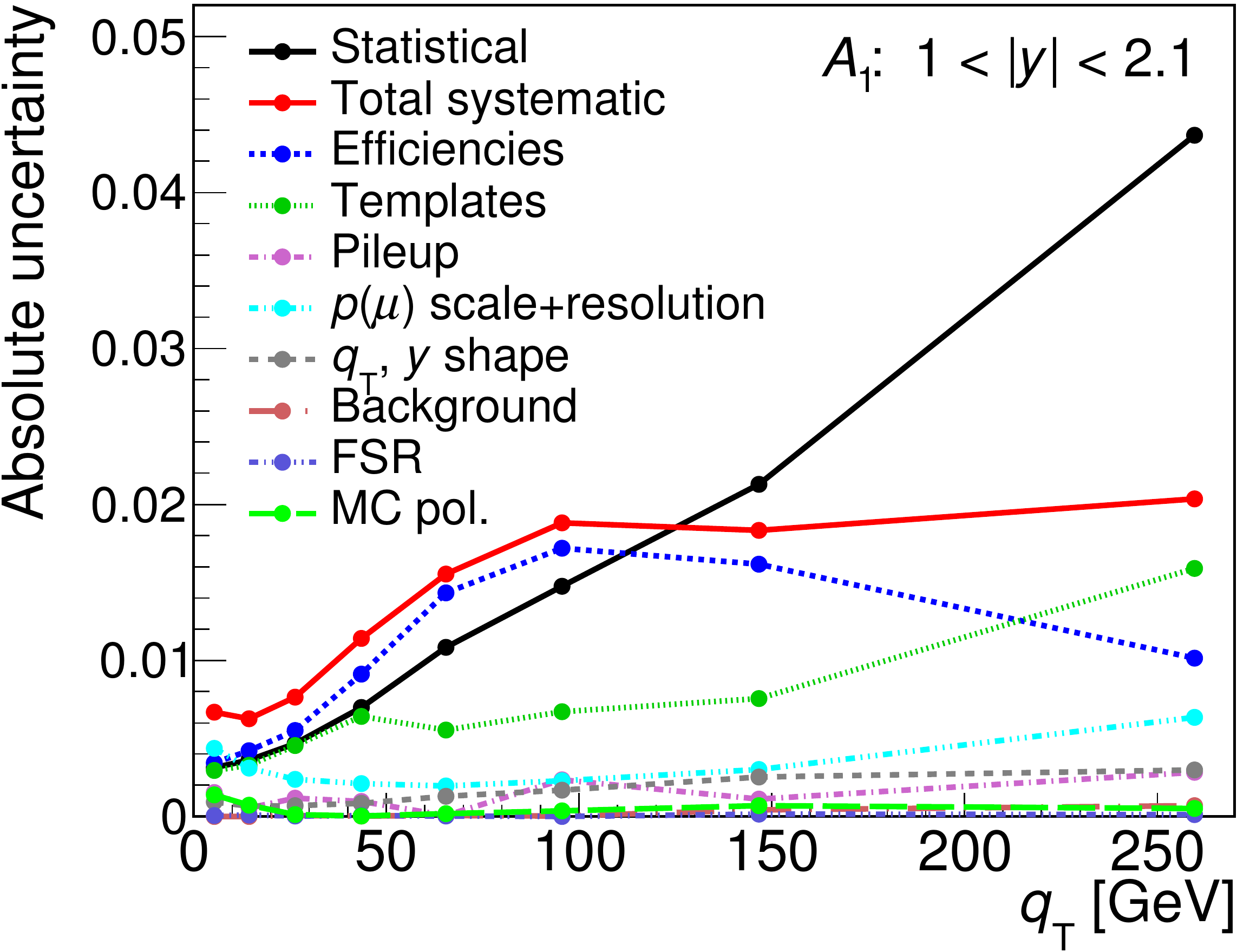}\\
\includegraphics[width=0.36\textwidth]{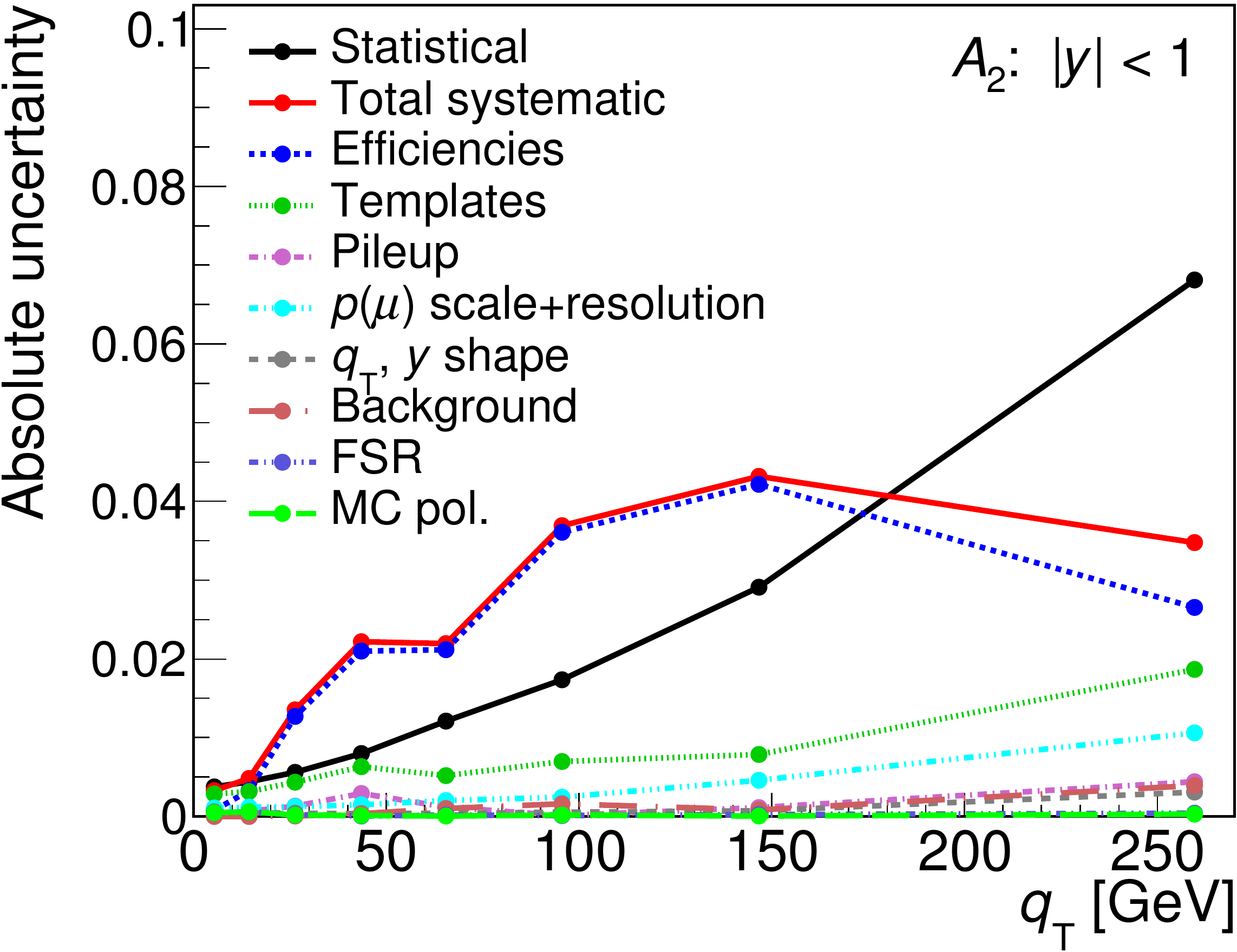}
\includegraphics[width=0.36\textwidth]{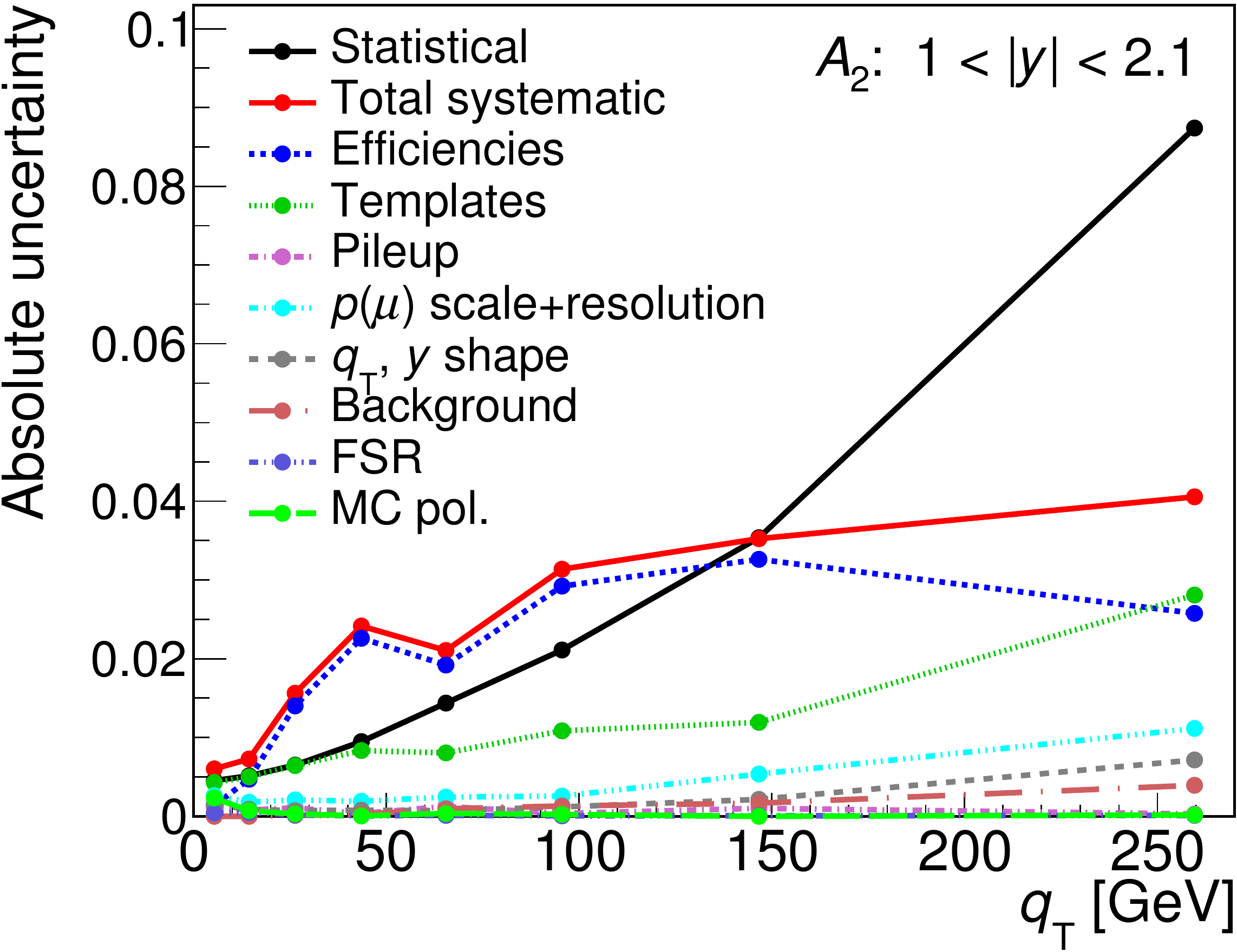}\\
\includegraphics[width=0.36\textwidth]{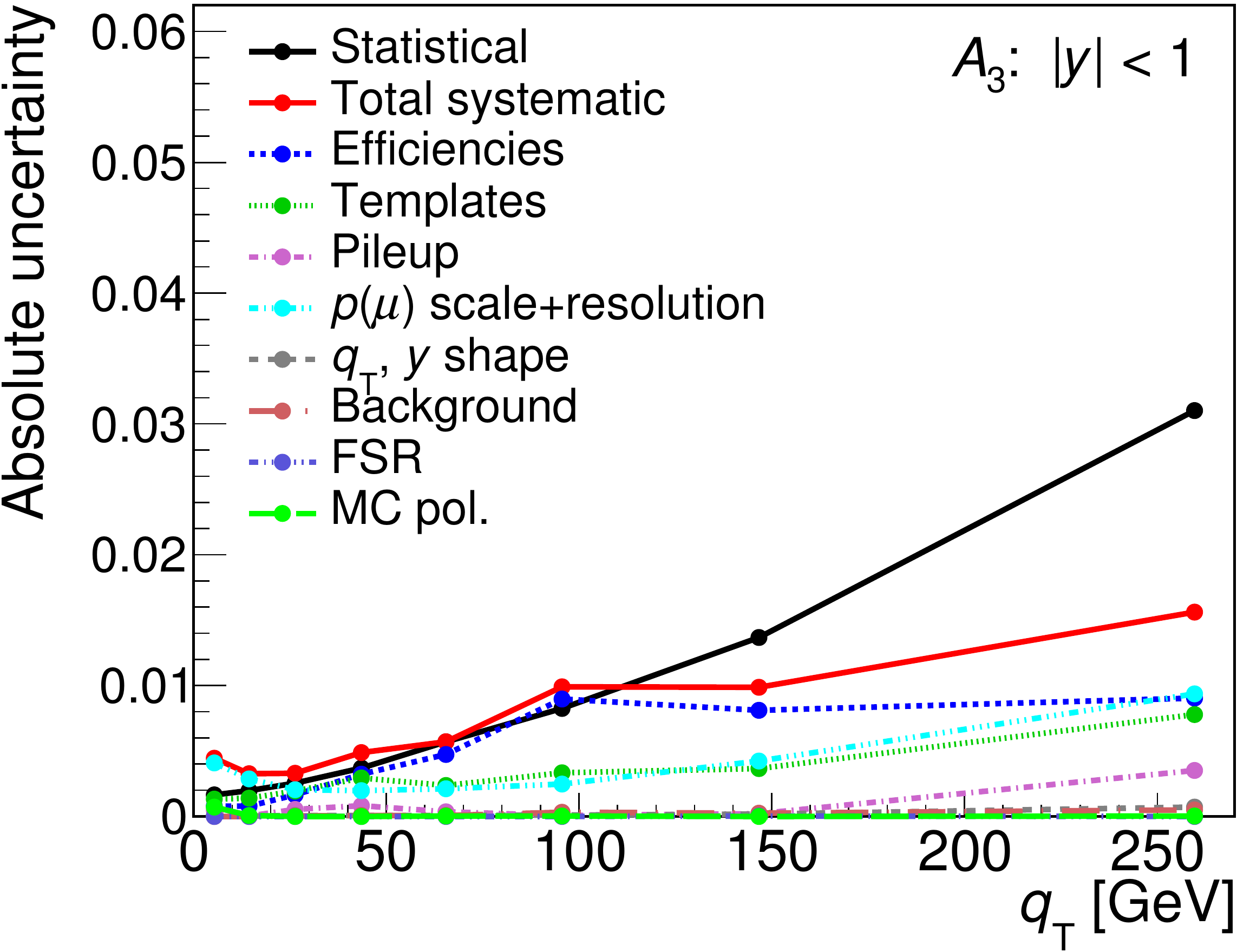}
\includegraphics[width=0.36\textwidth]{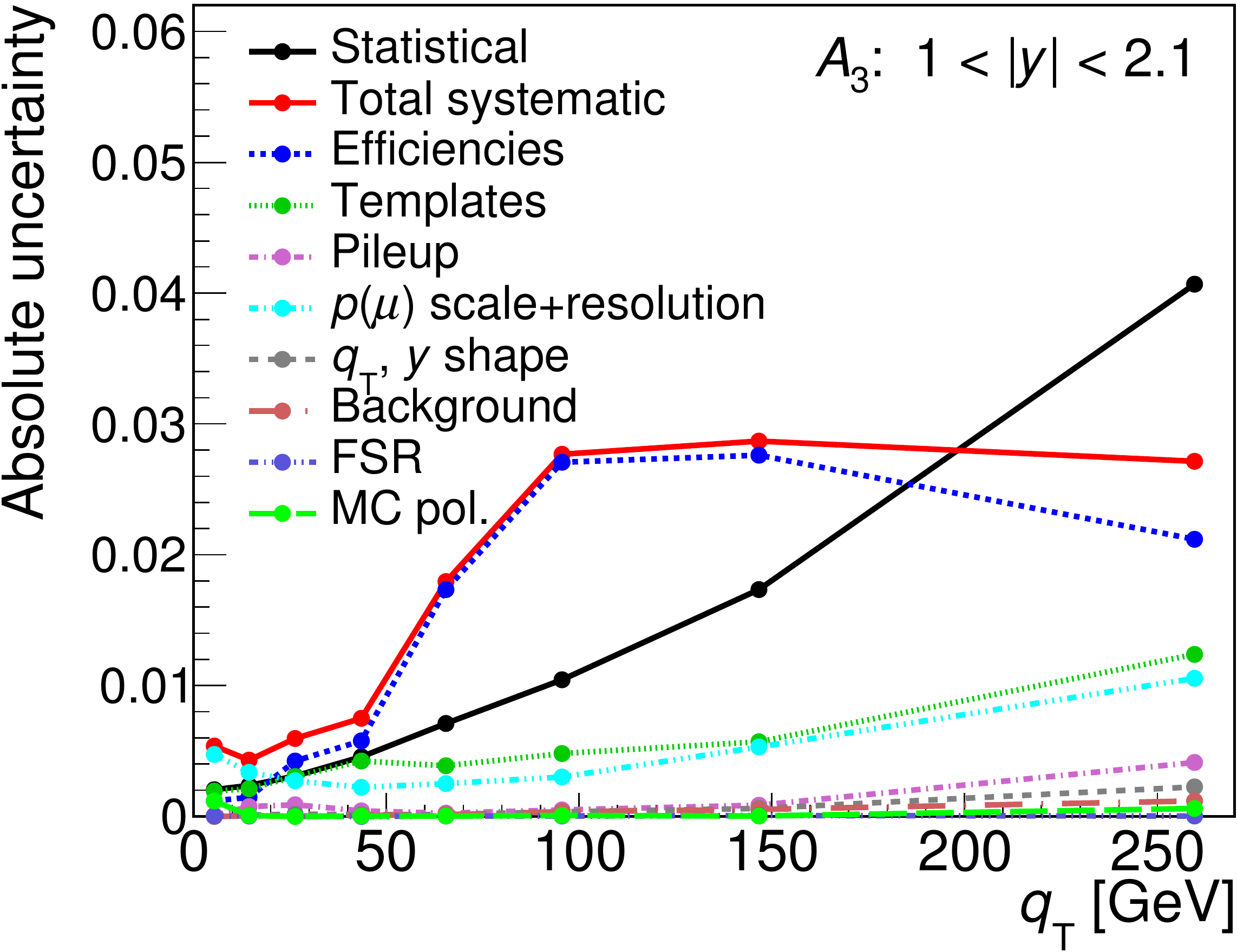}\\
\includegraphics[width=0.36\textwidth]{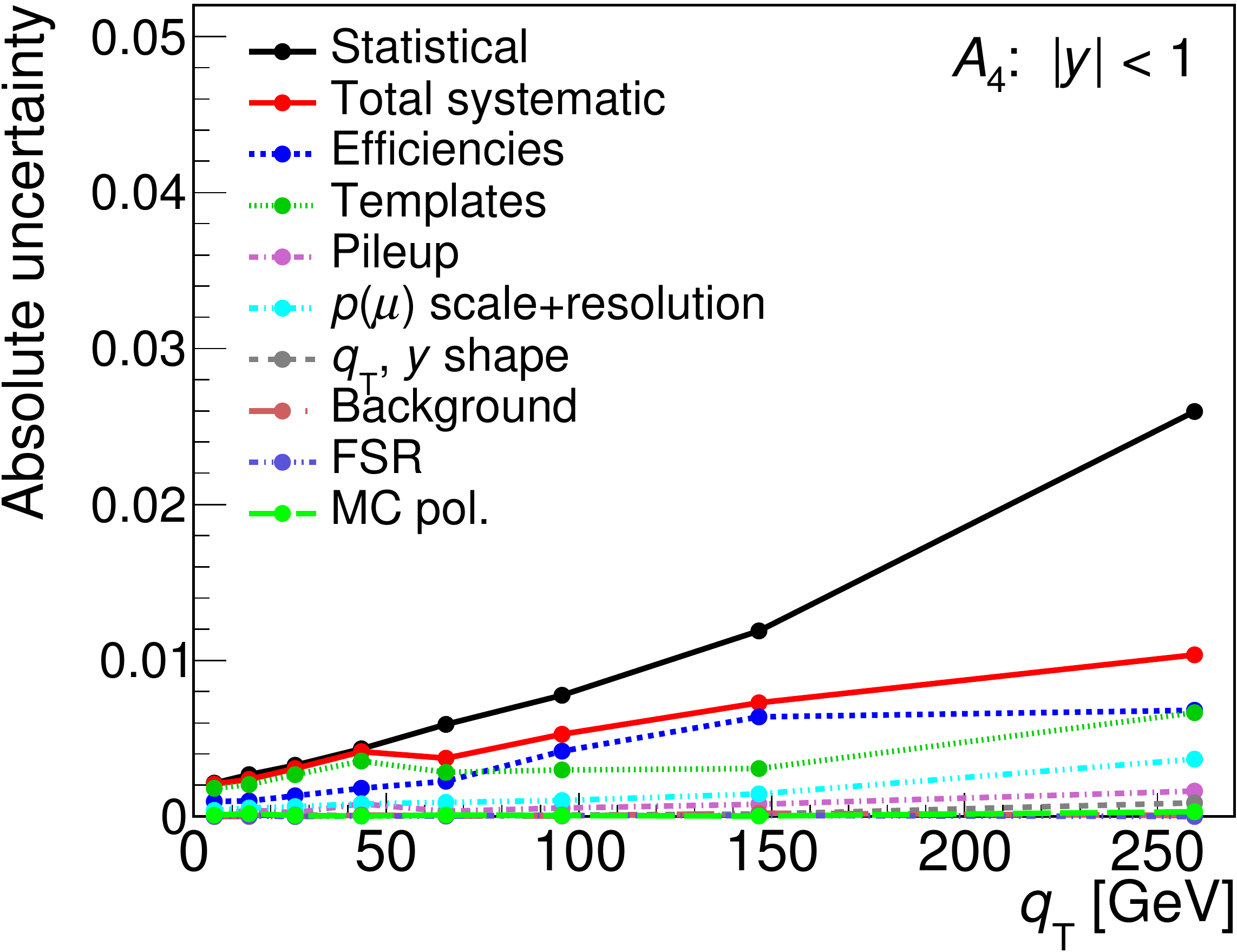}
\includegraphics[width=0.36\textwidth]{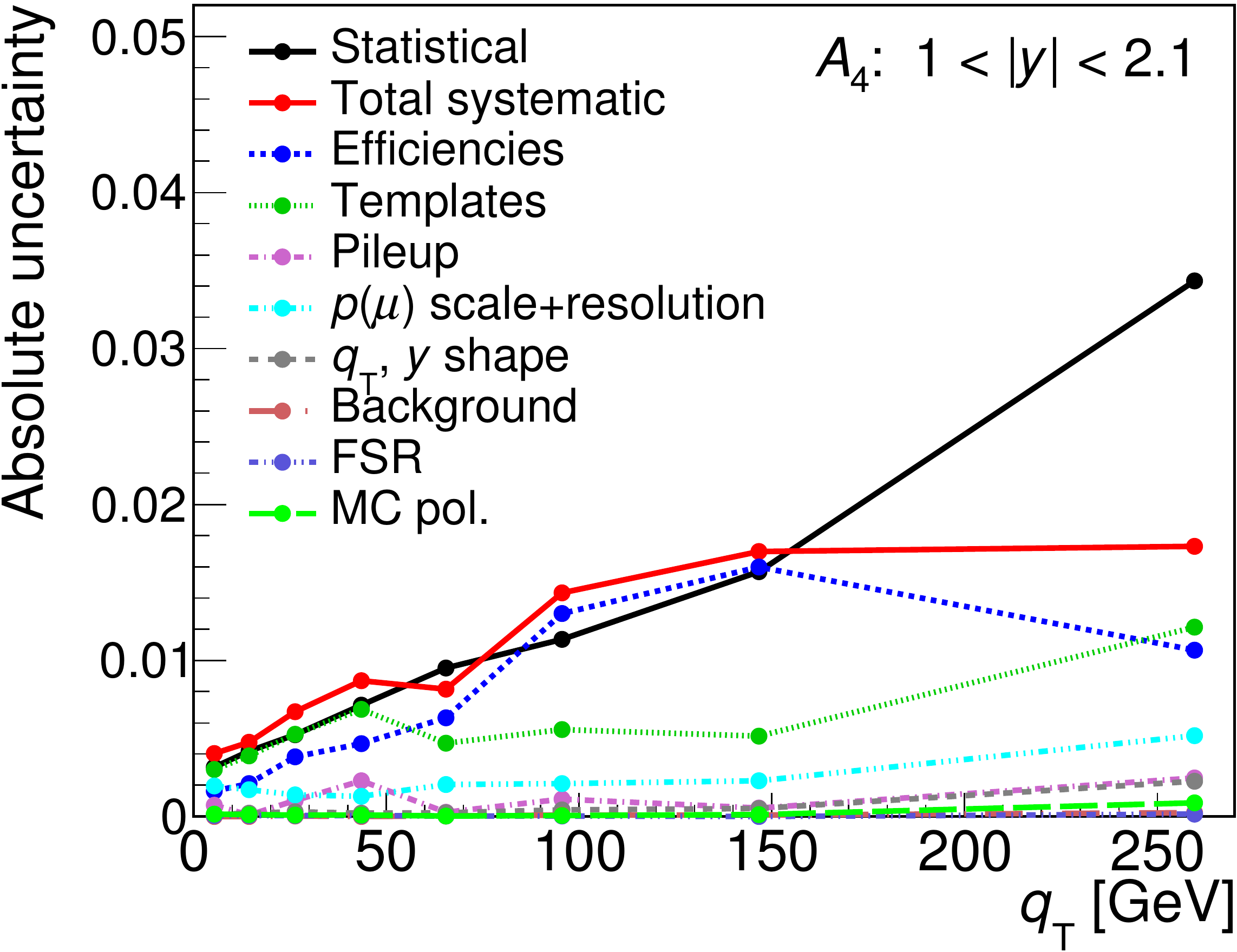}\\

\caption{Absolute uncertainties in the five angular coefficients $A_0$ to $A_4$. Each figure shows the $\qt$ dependence in the indicated ranges of $\abs{y}$.}
\label{fig:uncert}
\end{figure*}

\begin{figure*}[!ht]
 \centering
\includegraphics[width=0.329\textwidth] {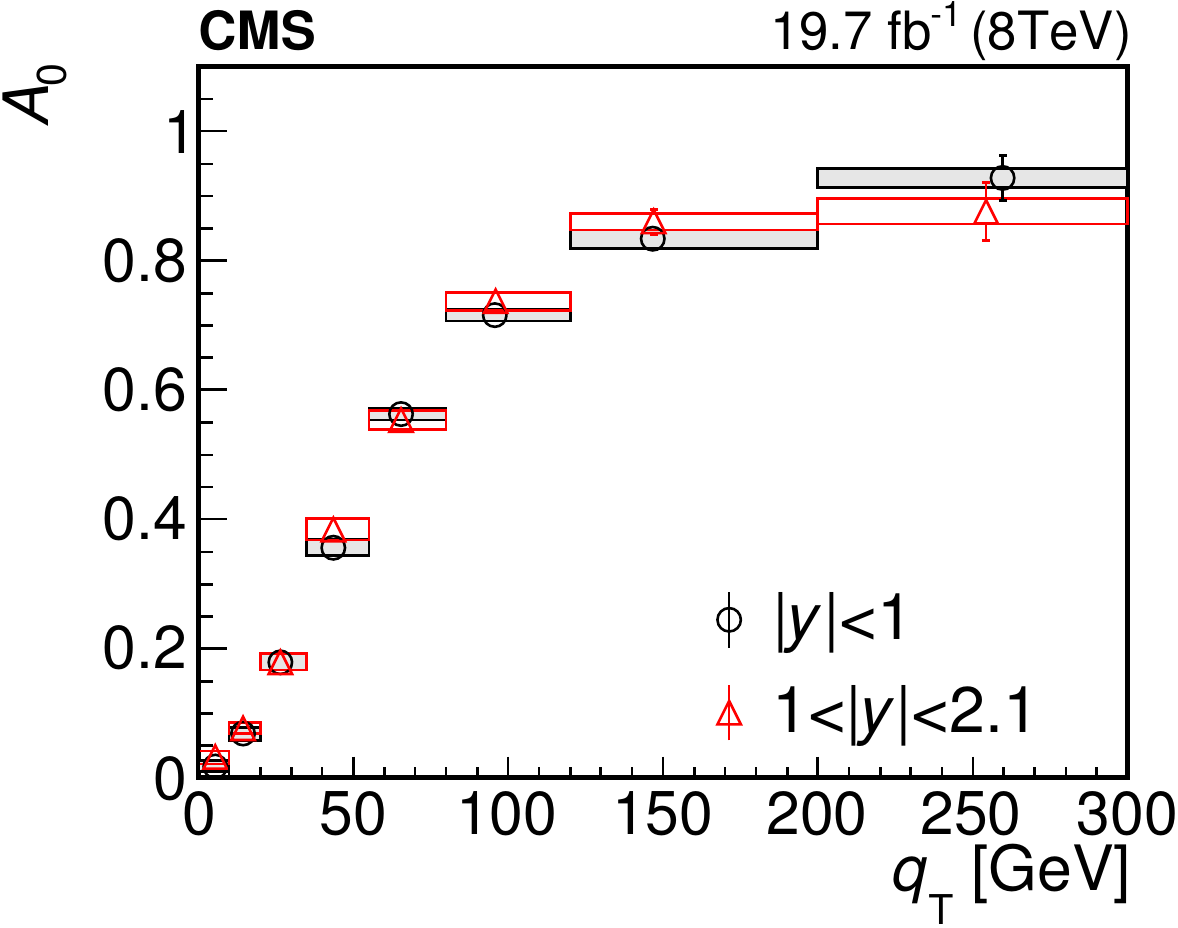}
\includegraphics[width=0.329\textwidth] {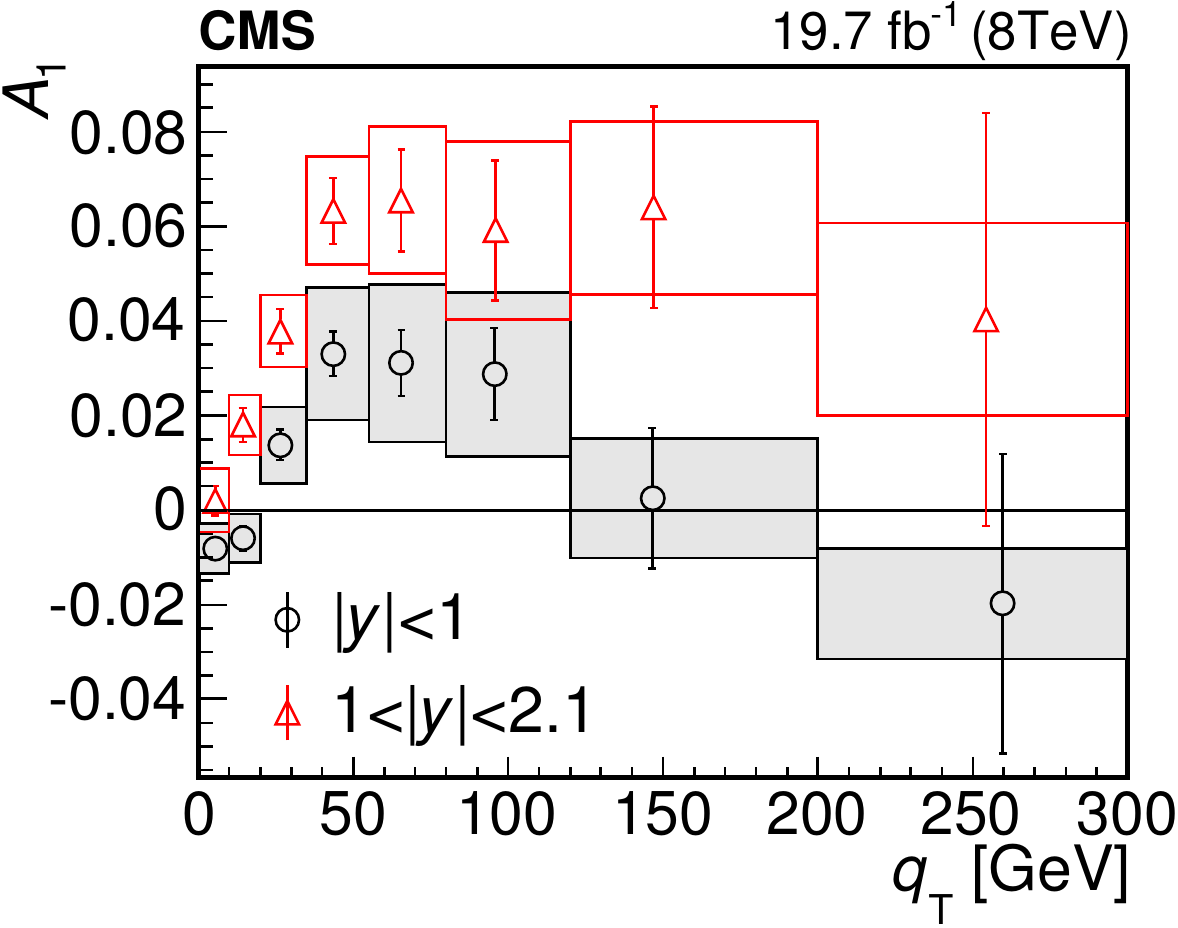}
\includegraphics[width=0.329\textwidth] {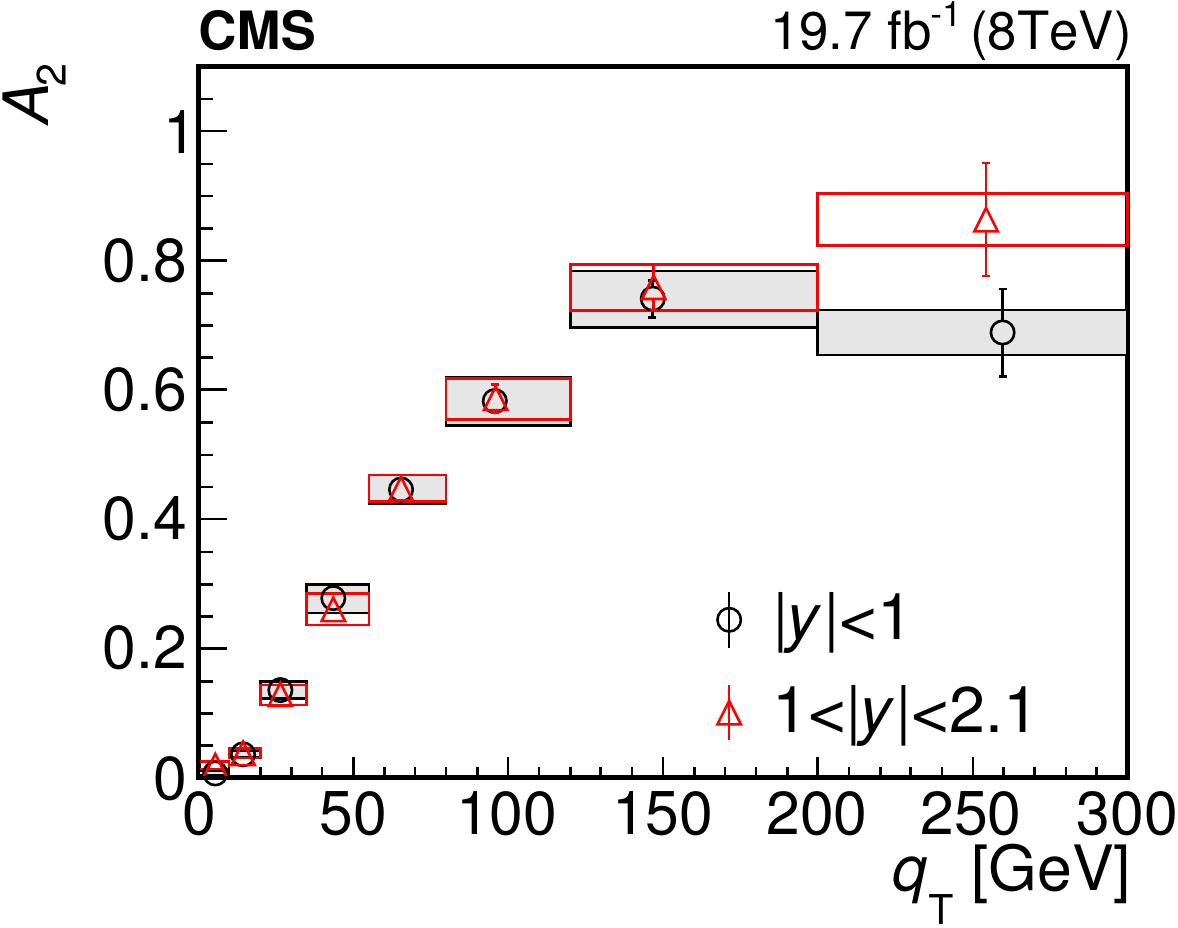}
\includegraphics[width=0.329\textwidth] {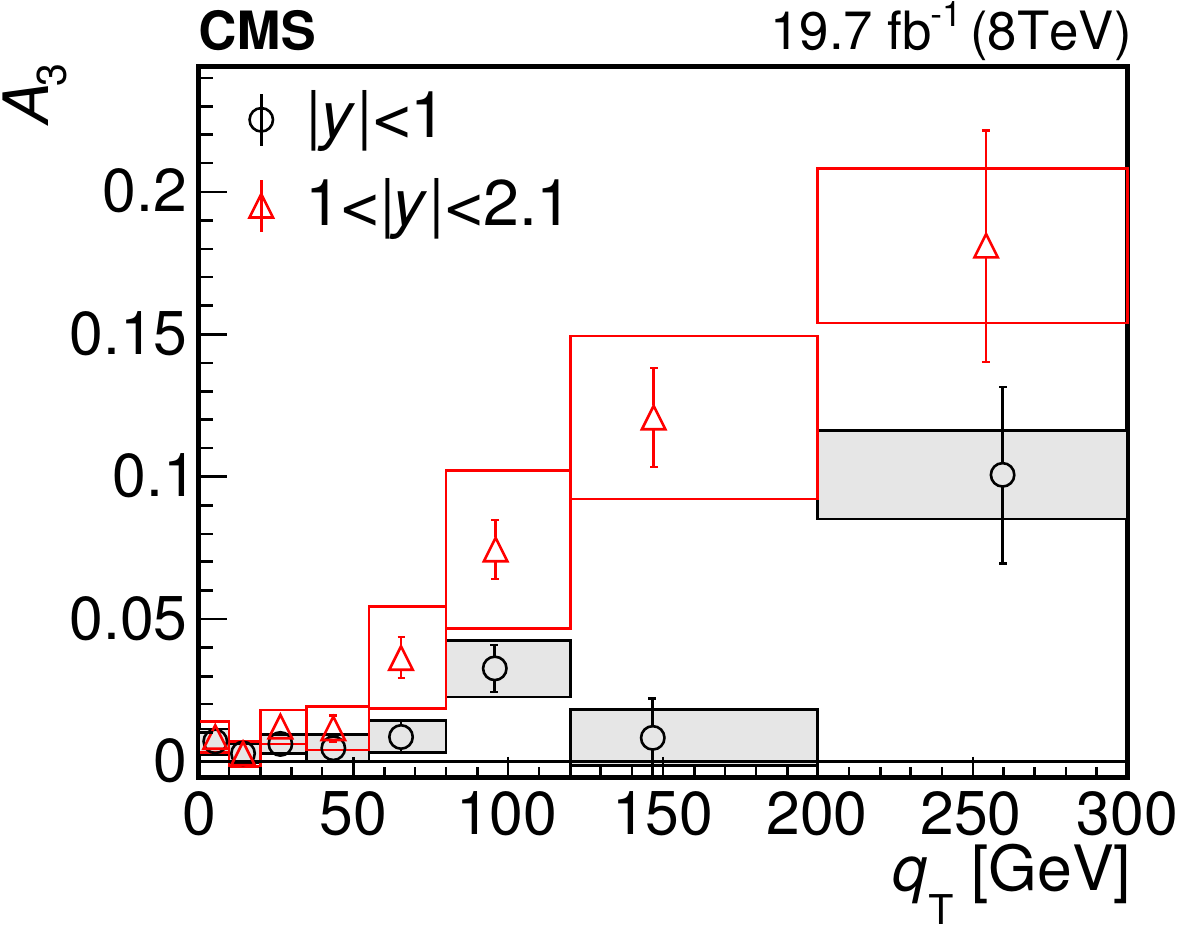}
\includegraphics[width=0.329\textwidth] {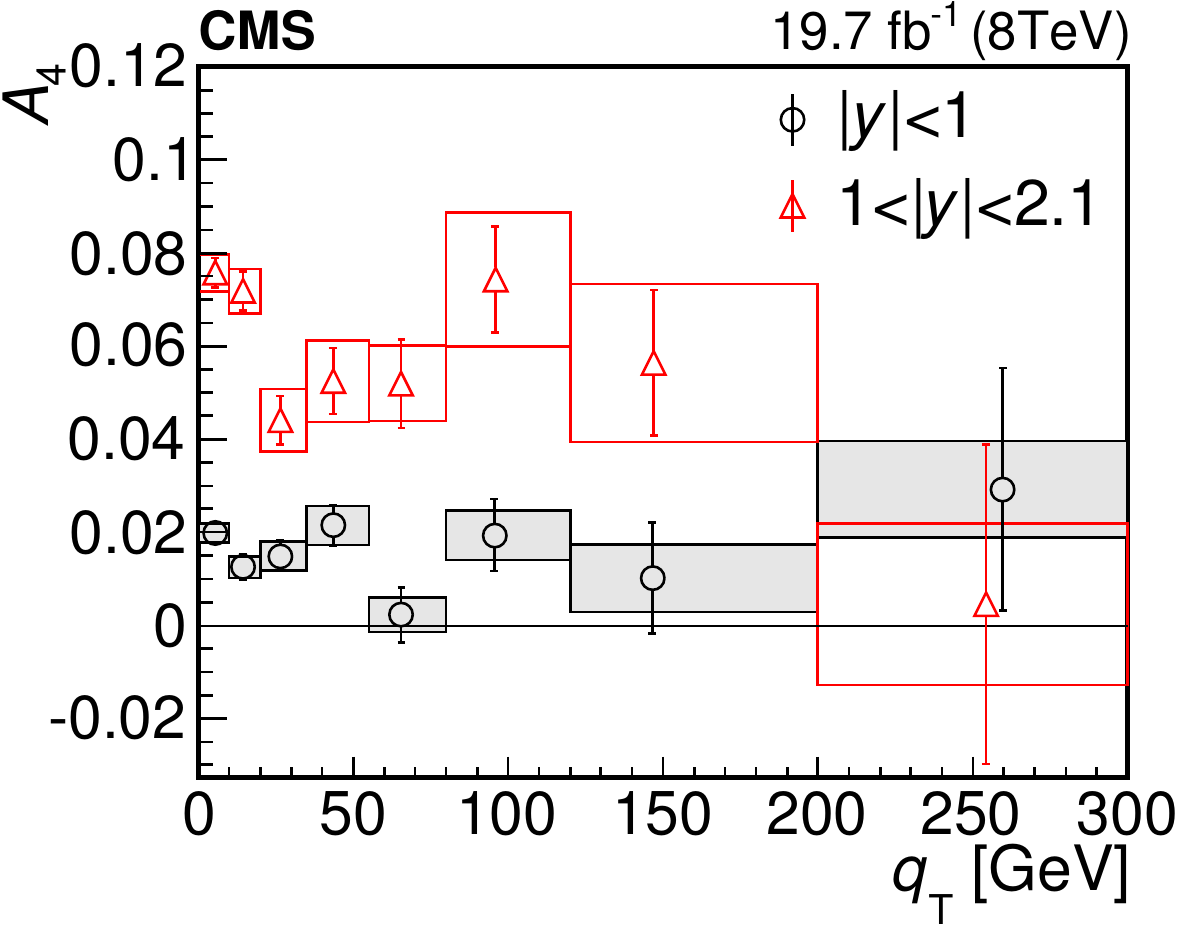}
\includegraphics[width=0.329\textwidth] {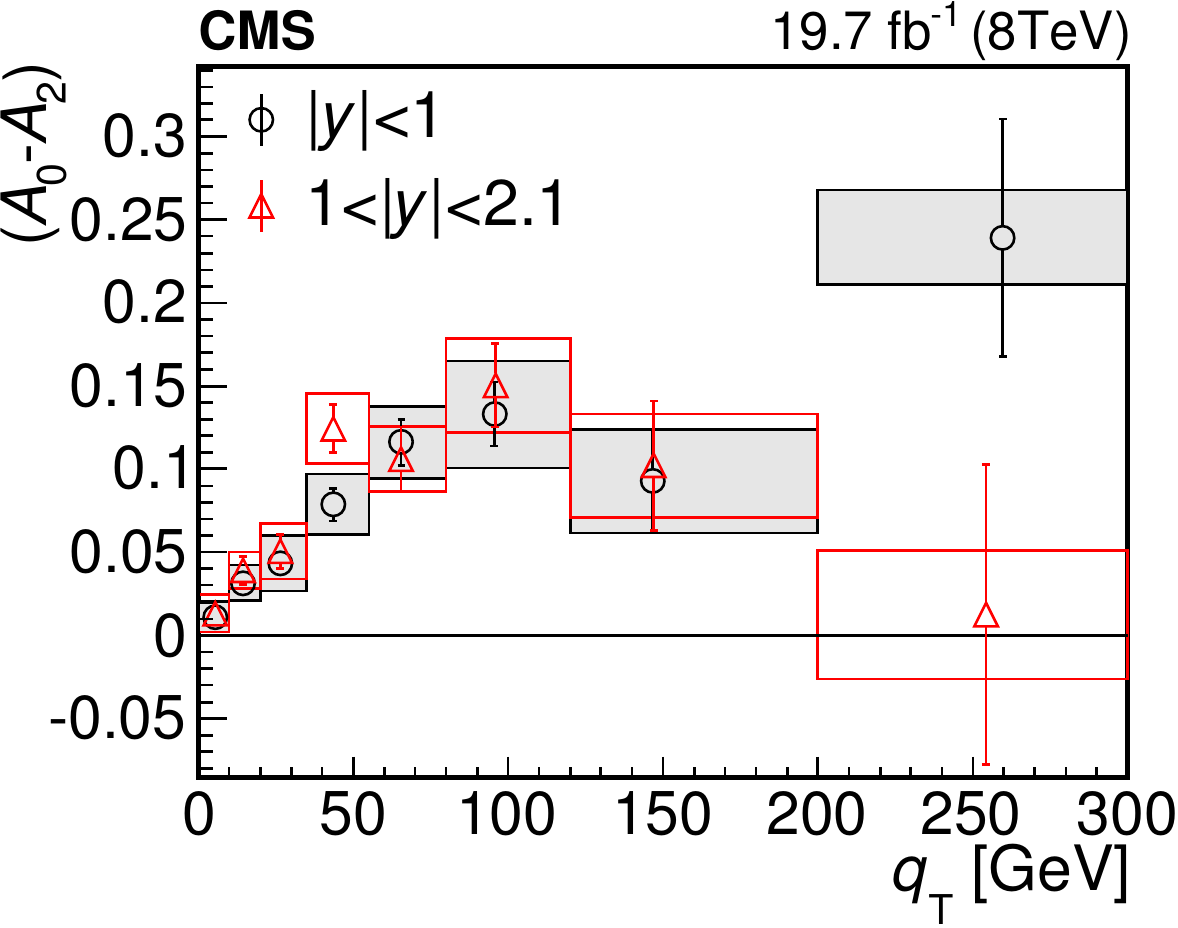}

\caption{Comparison of the five angular coefficients $A_i$ and $A_0-A_2$ measured in the Collins--Soper frame in bins of $\qt$ between $\abs{y}<1$ (circles) and $1<\abs{y}<2.1$ (triangles). The vertical bars represent the statistical uncertainties and the boxes the systematic uncertainties of the measurement.}
\label{fig:pol_res_cs_Ydep}
\end{figure*}

The results of the $\qt$ and $\abs{y}$ dependent
measurements of the angular coefficients $A_0$ to $A_4$ as well as the difference $A_0-A_2$ are presented along with
\MADGRAPH, \POWHEG, and \FEWZ (at NNLO) calculations in Figs.~\ref{fig:pol_res_cs_Y0} and \ref{fig:pol_res_cs_Y1}.
The various systematic uncertainties of the five angular coefficients $A_0$ to $A_4$ are presented in Fig.~\ref{fig:uncert}.
The values and uncertainties of the coefficients are provided in Tables~\ref{tab:res_y0} and \ref{tab:res_y1}.
The PDF sets used in the calculations are CTEQ6L for \MADGRAPH and CT10 for \POWHEG (at NLO) and \FEWZ (at NNLO).
The \MADGRAPH predictions for $A_4$ are systematically higher than those of \POWHEG and \FEWZ because \MADGRAPH uses a weak mixing angle calculated without considering radiative corrections.
The measured $A_0$ and $A_2$ coefficients agree better with the prediction of \MADGRAPH than with those of \POWHEG and \FEWZ, especially at high $\qt$.
 At $\qt=0$, the \POWHEG prediction
for $A_0$ is negative, which is unphysical and has been traced to approximations in the shower matching algorithm.
The \FEWZ prediction is shown for $\qt > 20$\GeV, where the calculations are considered reliable.
 We find that $A_0$($\qt$) and $A_2$($\qt$) are larger in pp collisions than those measured in  $\Pp\Pap$ collisions at the Tevatron.
The larger contribution from the qg process in pp collisions at the LHC is responsible for this difference.
We observe the violation of the Lam--Tung relation ($A_0=A_2$) anticipated by QCD calculations beyond leading order~\cite{Mirkes:1994dp}.
We find that $A_0>A_2$, especially for high $\qt$. In addition, we measure nonzero values of $A_1$ and $A_3$.
The comparison of the results for $\abs{y}<1$ and $1<\abs{y}<2.1$ is shown in Fig.~\ref{fig:pol_res_cs_Ydep}.

 In summary, we presented the five major angular coefficients, $A_{0}$ through $A_{4}$, for the production of the Z boson decaying to muon pairs as a function of $\qt$ and $\abs{y}$ in pp collisions. These results play an important role in future high-precision measurements, such as the measurement of the mass of the W boson and of the electroweak mixing angle.
Some theoretical predictions deviate from the measurements in $\qt$.
Further refinements of the theory are needed to achieve a better agreement with the experimental results.

\begin{table*}[p]
\centering
\topcaption{The five angular coefficients $A_0$ to $A_4$ and $A_0-A_2$ in bins of $\qt$ for $\abs{y}<1$.}\medskip
\label{tab:res_y0}
\setlength{\tabcolsep}{3.4pt}
 \renewcommand{\arraystretch}{1.2}
\begin{tabular}{c|rcc|rcc|rcc} \hline
$\qt$ [\GeVns{}] &$A_0$&$\pm\delta_{\text{stat}}$&$\pm\delta_{\text{syst}}$ &$A_1$&$\pm\delta_{\text{stat}}$&$\pm\delta_{\text{syst}}$&$A_2$&$\pm\delta_{\text{stat}}$&$\pm\delta_{\text{syst}}$\\ \hline
0--10       & $ 0.018$&$\pm  0.003$&$\pm  0.009$ & $-0.008$&$\pm  0.002$&$\pm  0.005$ & $ 0.007$&$\pm  0.004$&$\pm  0.003$ \\
10--20      & $ 0.068$&$\pm  0.004$&$\pm  0.010$ & $-0.006$&$\pm  0.003$&$\pm  0.005$ & $ 0.037$&$\pm  0.004$&$\pm  0.005$ \\
20--35      & $ 0.179$&$\pm  0.004$&$\pm  0.013$ & $ 0.014$&$\pm  0.003$&$\pm  0.008$ & $ 0.136$&$\pm  0.006$&$\pm  0.014$ \\
35--55      & $ 0.357$&$\pm  0.006$&$\pm  0.013$ & $ 0.033$&$\pm  0.005$&$\pm  0.014$ & $ 0.278$&$\pm  0.008$&$\pm  0.022$ \\
55--80      & $ 0.563$&$\pm  0.007$&$\pm  0.010$ & $ 0.031$&$\pm  0.007$&$\pm  0.017$ & $ 0.447$&$\pm  0.012$&$\pm  0.022$ \\
80--120     & $ 0.716$&$\pm  0.010$&$\pm  0.009$ & $ 0.029$&$\pm  0.010$&$\pm  0.017$ & $ 0.583$&$\pm  0.017$&$\pm  0.037$ \\
120--200    & $ 0.834$&$\pm  0.015$&$\pm  0.014$ & $ 0.002$&$\pm  0.015$&$\pm  0.013$ & $ 0.741$&$\pm  0.029$&$\pm  0.043$ \\
$>$200        & $ 0.928$&$\pm  0.035$&$\pm  0.015$ & $-0.020$&$\pm  0.032$&$\pm  0.012$ & $ 0.689$&$\pm  0.068$&$\pm  0.035$ \\  \hline

$\qt$ [\GeVns{}] &$A_3$&$\pm\delta_{\text{stat}}$&$\pm\delta_{\text{syst}}$&$A_4$&$\pm\delta_{\text{stat}}$&$\pm\delta_{\text{syst}}$&$A_0-A_2$&$\pm\delta_{\text{stat}}$&$\pm\delta_{\text{syst}}$\\ \hline
0--10       & $ 0.007$&$\pm  0.002$&$\pm  0.004$ & $ 0.020$&$\pm  0.002$&$\pm  0.002$ &  $ 0.011$&$\pm  0.005$&$\pm  0.009$ \\
10--20      & $ 0.003$&$\pm  0.002$&$\pm  0.003$ & $ 0.013$&$\pm  0.003$&$\pm  0.002$ &  $ 0.032$&$\pm  0.006$&$\pm  0.011$ \\
20--35      & $ 0.006$&$\pm  0.003$&$\pm  0.003$ & $ 0.015$&$\pm  0.003$&$\pm  0.003$ &  $ 0.043$&$\pm  0.007$&$\pm  0.016$ \\
35--55      & $ 0.005$&$\pm  0.004$&$\pm  0.005$ & $ 0.021$&$\pm  0.004$&$\pm  0.004$ &  $ 0.079$&$\pm  0.010$&$\pm  0.018$ \\
55--80      & $ 0.009$&$\pm  0.006$&$\pm  0.006$ & $ 0.002$&$\pm  0.006$&$\pm  0.004$ &  $ 0.116$&$\pm  0.014$&$\pm  0.022$ \\
80--120     & $ 0.033$&$\pm  0.008$&$\pm  0.010$ & $ 0.019$&$\pm  0.008$&$\pm  0.005$ &  $ 0.133$&$\pm  0.019$&$\pm  0.032$ \\
120--200    & $ 0.008$&$\pm  0.014$&$\pm  0.010$ & $ 0.010$&$\pm  0.012$&$\pm  0.007$ &  $ 0.093$&$\pm  0.031$&$\pm  0.031$ \\
$>$200        & $ 0.101$&$\pm  0.031$&$\pm  0.016$ & $ 0.029$&$\pm  0.026$&$\pm  0.010$ &  $ 0.239$&$\pm  0.072$&$\pm  0.028$ \\ \hline
\end{tabular}
\end{table*}
\begin{table*}[p]
\centering
\topcaption{The five angular coefficients $A_0$ to $A_4$ and $A_0-A_2$ in bins of $\qt$ for $1<\abs{y}<2.1$.}\medskip
\label{tab:res_y1}
\setlength{\tabcolsep}{3.4pt}
\renewcommand{\arraystretch}{1.2}
\begin{tabular}{c|rcc|rcc|rcc} \hline
$\qt$ [\GeVns{}] &$A_0$&$\pm\delta_{\text{stat}}$&$\pm\delta_{\text{syst}}$ &$A_1$&$\pm\delta_{\text{stat}}$&$\pm\delta_{\text{syst}}$&$A_2$&$\pm\delta_{\text{stat}}$&$\pm\delta_{\text{syst}}$\\ \hline
0--10       & $ 0.032$&$\pm  0.005$&$\pm  0.010$ & $ 0.002$&$\pm  0.003$&$\pm  0.007$ & $ 0.019$&$\pm  0.005$&$\pm  0.006$\\
10--20      & $ 0.077$&$\pm  0.006$&$\pm  0.009$ & $ 0.018$&$\pm  0.004$&$\pm  0.006$ & $ 0.038$&$\pm  0.005$&$\pm  0.007$\\
20--35      & $ 0.179$&$\pm  0.008$&$\pm  0.013$ & $ 0.038$&$\pm  0.005$&$\pm  0.008$ & $ 0.129$&$\pm  0.006$&$\pm  0.016$\\
35--55      & $ 0.385$&$\pm  0.011$&$\pm  0.017$ & $ 0.063$&$\pm  0.007$&$\pm  0.011$ & $ 0.260$&$\pm  0.009$&$\pm  0.024$\\
55--80      & $ 0.554$&$\pm  0.013$&$\pm  0.015$ & $ 0.066$&$\pm  0.011$&$\pm  0.016$ & $ 0.448$&$\pm  0.014$&$\pm  0.021$\\
80--120     & $ 0.737$&$\pm  0.015$&$\pm  0.014$ & $ 0.059$&$\pm  0.015$&$\pm  0.019$ & $ 0.587$&$\pm  0.021$&$\pm  0.031$\\
120--200    & $ 0.860$&$\pm  0.020$&$\pm  0.012$ & $ 0.064$&$\pm  0.021$&$\pm  0.018$ & $ 0.758$&$\pm  0.035$&$\pm  0.035$\\
$>$200        & $ 0.876$&$\pm  0.045$&$\pm  0.020$ & $ 0.040$&$\pm  0.044$&$\pm  0.020$ & $ 0.864$&$\pm  0.087$&$\pm  0.041$\\ \hline
$\qt$ [\GeVns{}] &$A_3$&$\pm\delta_{\text{stat}}$&$\pm\delta_{\text{syst}}$&$A_4$&$\pm\delta_{\text{stat}}$&$\pm\delta_{\text{syst}}$&$A_0-A_2$&$\pm\delta_{\text{stat}}$&$\pm\delta_{\text{syst}}$\\ \hline
0--10       & $ 0.009$&$\pm  0.002$&$\pm  0.005$ & $ 0.076$&$\pm  0.003$&$\pm  0.004$ & $ 0.013$&$\pm  0.007$&$\pm  0.011$\\
10--20      & $ 0.003$&$\pm  0.002$&$\pm  0.004$ & $ 0.072$&$\pm  0.004$&$\pm  0.005$ & $ 0.039$&$\pm  0.008$&$\pm  0.011$\\
20--35      & $ 0.012$&$\pm  0.003$&$\pm  0.006$ & $ 0.044$&$\pm  0.005$&$\pm  0.007$ & $ 0.051$&$\pm  0.010$&$\pm  0.017$\\
35--55      & $ 0.012$&$\pm  0.005$&$\pm  0.008$ & $ 0.052$&$\pm  0.007$&$\pm  0.009$ & $ 0.124$&$\pm  0.014$&$\pm  0.021$\\
55--80      & $ 0.036$&$\pm  0.007$&$\pm  0.018$ & $ 0.052$&$\pm  0.009$&$\pm  0.008$ & $ 0.106$&$\pm  0.019$&$\pm  0.019$\\
80--120     & $ 0.074$&$\pm  0.010$&$\pm  0.028$ & $ 0.074$&$\pm  0.011$&$\pm  0.014$ & $ 0.150$&$\pm  0.025$&$\pm  0.028$\\
120--200    & $ 0.121$&$\pm  0.017$&$\pm  0.029$ & $ 0.056$&$\pm  0.016$&$\pm  0.017$ & $ 0.102$&$\pm  0.039$&$\pm  0.031$\\
$>$200        & $ 0.181$&$\pm  0.041$&$\pm  0.027$ & $ 0.005$&$\pm  0.034$&$\pm  0.017$ & $ 0.012$&$\pm  0.090$&$\pm  0.039$\\ \hline
\end{tabular}
\end{table*}

\begin{acknowledgments}
We congratulate our colleagues in the CERN accelerator departments for the excellent performance of the LHC and thank the technical and administrative staffs at CERN and at other CMS institutes for their contributions to the success of the CMS effort. In addition, we gratefully acknowledge the computing centers and personnel of the Worldwide LHC Computing Grid for delivering so effectively the computing infrastructure essential to our analyses. Finally, we acknowledge the enduring support for the construction and operation of the LHC and the CMS detector provided by the following funding agencies: BMWFW and FWF (Austria); FNRS and FWO (Belgium); CNPq, CAPES, FAPERJ, and FAPESP (Brazil); MES (Bulgaria); CERN; CAS, MoST, and NSFC (China); COLCIENCIAS (Colombia); MSES and CSF (Croatia); RPF (Cyprus); MoER, ERC IUT and ERDF (Estonia); Academy of Finland, MEC, and HIP (Finland); CEA and CNRS/IN2P3 (France); BMBF, DFG, and HGF (Germany); GSRT (Greece); OTKA and NIH (Hungary); DAE and DST (India); IPM (Iran); SFI (Ireland); INFN (Italy); MSIP and NRF (Republic of Korea); LAS (Lithuania); MOE and UM (Malaysia); CINVESTAV, CONACYT, SEP, and UASLP-FAI (Mexico); MBIE (New Zealand); PAEC (Pakistan); MSHE and NSC (Poland); FCT (Portugal); JINR (Dubna); MON, RosAtom, RAS and RFBR (Russia); MESTD (Serbia); SEIDI and CPAN (Spain); Swiss Funding Agencies (Switzerland); MST (Taipei); ThEPCenter, IPST, STAR and NSTDA (Thailand); TUBITAK and TAEK (Turkey); NASU and SFFR (Ukraine); STFC (United Kingdom); DOE and NSF (USA).

Individuals have received support from the Marie-Curie program and the European Research Council and EPLANET (European Union); the Leventis Foundation; the A. P. Sloan Foundation; the Alexander von Humboldt Foundation; the Belgian Federal Science Policy Office; the Fonds pour la Formation \`a la Recherche dans l'Industrie et dans l'Agriculture (FRIA-Belgium); the Agentschap voor Innovatie door Wetenschap en Technologie (IWT-Belgium); the Ministry of Education, Youth and Sports (MEYS) of the Czech Republic; the Council of Science and Industrial Research, India; the HOMING PLUS program of the Foundation for Polish Science, cofinanced from European Union, Regional Development Fund; the Compagnia di San Paolo (Torino); the Consorzio per la Fisica (Trieste); MIUR project 20108T4XTM (Italy); the Thalis and Aristeia programs cofinanced by EU-ESF and the Greek NSRF; and the National Priorities Research Program by Qatar National Research Fund.
\end{acknowledgments}

\bibliography{auto_generated}

\providecommand{\href}[2]{#2}\begingroup\raggedright\begin{thebibliography}{10}%
\makeatletter
\providecommand{\hrefCMSnoop }[0]{\@secondoftwo}%
\makeatother
\providecommand{\doi}{\texttt{doi:}\begingroup \urlstyle{tt}\Url}

\bibitem{Collins:1977iv}
\hrefCMSnoop {}{J.~C. Collins and D.~E. Soper, ``Angular distribution of
  dileptons in high-energy hadron collisions'',} \textit{ Phys. Rev. D}
  \textbf{ 16} (1977) 2219,
\href{http://dx.doi.org/10.1103/PhysRevD.16.2219}{\doi{10.1103/PhysRevD.16.2219}}.

\bibitem{Aaltonen:2011nr}
\hrefCMSnoop {}{{CDF} Collaboration, ``First Measurement of the Angular
  Coefficients of {Drell-Yan} $\rm{e^{+}e^{-}}$ pairs in the {Z} Mass Region
  from $\text{p}\bar{\text{p}}$ Collisions at $\sqrt{s}$ = 1.96 {TeV}'',}
  \textit{ Phys. Rev. Lett.} \textbf{ 106} (2011) 241801,
  \href{http://dx.doi.org/10.1103/PhysRevLett.106.241801}{\doi{10.1103/PhysRevLett.106.241801}},
\href{http://www.arXiv.org/abs/1103.5699}{\texttt{arXiv:1103.5699}}.

\bibitem{Nason:2004rx}
\hrefCMSnoop {}{P.~Nason, ``A new method for combining {NLO} {QCD} with shower
  Monte Carlo algorithms'',} \textit{ JHEP} \textbf{ 11} (2004) 040,
  \href{http://dx.doi.org/10.1088/1126-6708/2004/11/040}{\doi{10.1088/1126-6708/2004/11/040}},
\href{http://www.arXiv.org/abs/hep-ph/0409146}{\texttt{arXiv:hep-ph/0409146}}.

\bibitem{Frixione:2007vw}
\hrefCMSnoop {}{S.~Frixione, P.~Nason, and C.~Oleari, ``Matching {NLO} {QCD}
  computations with parton shower simulations: the {POWHEG} method'',} \textit{
  JHEP} \textbf{ 11} (2007) 070,
  \href{http://dx.doi.org/10.1088/1126-6708/2007/11/070}{\doi{10.1088/1126-6708/2007/11/070}},
\href{http://www.arXiv.org/abs/0709.2092}{\texttt{arXiv:0709.2092}}.

\bibitem{Alioli:2010xd}
\hrefCMSnoop {}{S.~Alioli, P.~Nason, C.~Oleari, and E.~Re, ``A general
  framework for implementing {NLO} calculations in shower Monte Carlo programs:
  the {POWHEG} {BOX}'',} \textit{ JHEP} \textbf{ 06} (2010) 043,
  \href{http://dx.doi.org/10.1007/JHEP06(2010)043}{\doi{10.1007/JHEP06(2010)043}},
\href{http://www.arXiv.org/abs/1002.2581}{\texttt{arXiv:1002.2581}}.

\bibitem{Alioli:2008gx}
\hrefCMSnoop {}{S.~Alioli, P.~Nason, C.~Oleari, and E.~Re, ``{NLO} vector-boson
  production matched with shower in {POWHEG}'',} \textit{ JHEP} \textbf{ 07}
  (2008) 060,
  \href{http://dx.doi.org/10.1088/1126-6708/2008/07/060}{\doi{10.1088/1126-6708/2008/07/060}},
\href{http://www.arXiv.org/abs/0805.4802}{\texttt{arXiv:0805.4802}}.

\bibitem{qq}
\hrefCMSnoop {}{J.~C. Collins, ``Simple Prediction of Quantum Chromodynamics
  for Angular Distribution of Dileptons in Hadron Collisions'',} \textit{ Phys.
  Rev. Lett.} \textbf{ 42} (1979) 291,
  \href{http://dx.doi.org/10.1103/PhysRevLett.42.291}{\doi{10.1103/PhysRevLett.42.291}}.

\bibitem{Boer:2006eq}
\hrefCMSnoop {}{D.~Boer and W.~Vogelsang, ``{Drell-Yan} lepton angular
  distribution at small transverse momentum'',} \textit{ Phys. Rev. D} \textbf{
  74} (2006) 014004,
  \href{http://dx.doi.org/10.1103/PhysRevD.74.014004}{\doi{10.1103/PhysRevD.74.014004}},
\href{http://www.arXiv.org/abs/hep-ph/0604177}{\texttt{arXiv:hep-ph/0604177}}.

\bibitem{berger}
\hrefCMSnoop {}{E.~L. Berger, J.-W. Qiu, and R.~A. Rodriguez-Pedraza, ``Angular
  distribution of leptons from the decay of massive vector bosons'',} \textit{
  Phys. Lett. B} \textbf{ 656} (2007) 74,
  \href{http://dx.doi.org/10.1016/j.physletb.2007.09.008}{\doi{10.1016/j.physletb.2007.09.008}},
  \href{http://www.arXiv.org/abs/0707.3150}{\texttt{arXiv:0707.3150}}.

\bibitem{bodek}
\hrefCMSnoop {}{A.~Bodek, ``A simple event weighting technique for optimizing
  the measurement of the forward-backward asymmetry of Drell-Yan dilepton pairs
  at hadron colliders'',} \textit{ Eur. Phys. J. C} \textbf{ 67} (2010) 321,
  \href{http://dx.doi.org/10.1140/epjc/s10052-010-1287-5}{\doi{10.1140/epjc/s10052-010-1287-5}},
  \href{http://www.arXiv.org/abs/0911.2850}{\texttt{arXiv:0911.2850}}.

\bibitem{qg}
\hrefCMSnoop {}{J.~Lindfors, ``Angular Distribution of Large $q_{\rm T}$ Muon
  Pairs in Different Reference Frames'',} \textit{ Physica Scripta} \textbf{
  20} (1979) 19,
  \href{http://dx.doi.org/10.1088/0031-8949/20/1/003}{\doi{10.1088/0031-8949/20/1/003}}.

\bibitem{Lam:1978zr}
\hrefCMSnoop {}{C.~S. Lam and W.-K. Tung, ``Structure function relations at
  large transverse momenta in Lepton-pair production processes'',} \textit{
  Phys. Lett. B} \textbf{ 80} (1979) 228,
\href{http://dx.doi.org/10.1016/0370-2693(79)90204-1}{\doi{10.1016/0370-2693(79)90204-1}}.

\bibitem{Rinvariance}
\hrefCMSnoop {}{P.~Faccioli, C.~Lourenco, and J.~Seixas, ``Rotation-Invariant
  Relations in Vector Meson Decays into Fermion Pairs'',} \textit{ Phys. Rev.
  Lett.} \textbf{ 105} (2010) 061601,
  \href{http://dx.doi.org/10.1103/PhysRevLett.105.061601}{\doi{10.1103/PhysRevLett.105.061601}},
  \href{http://www.arXiv.org/abs/1005.2601}{\texttt{arXiv:1005.2601}}.

\bibitem{LT_ref}
\hrefCMSnoop {}{P.~Faccioli, C.~Lourenco, J.~Seixas, and H.~K. Woehri,
  ``Model-independent constraints on the shape parameters of dilepton angular
  distributions'',} \textit{ Phys. Rev. D} \textbf{ 83} (2011) 056008,
  \href{http://dx.doi.org/10.1103/PhysRevD.83.056008}{\doi{10.1103/PhysRevD.83.056008}},
  \href{http://www.arXiv.org/abs/1102.3946}{\texttt{arXiv:1102.3946}}.

\bibitem{Chatrchyan:2011ig}
\hrefCMSnoop {}{{CMS} Collaboration, ``Measurement of the Polarization of {W}
  Bosons with Large Transverse Momenta in {W+Jets} Events at the {LHC}'',}
  \textit{ Phys. Rev. Lett.} \textbf{ 107} (2011) 021802,
  \href{http://dx.doi.org/10.1103/PhysRevLett.107.021802}{\doi{10.1103/PhysRevLett.107.021802}},
\href{http://www.arXiv.org/abs/1104.3829}{\texttt{arXiv:1104.3829}}.

\bibitem{ATLAS:2012au}
\hrefCMSnoop {}{{ATLAS} Collaboration, ``Measurement of the polarisation of
  {$W$} bosons produced with large transverse momentum in $\rm{pp}$ collisions
  at $\sqrt{s}=7$ {TeV} with the {ATLAS} experiment'',} \textit{ Eur. Phys. J.
  C} \textbf{ 72} (2012) 2001,
  \href{http://dx.doi.org/10.1140/epjc/s10052-012-2001-6}{\doi{10.1140/epjc/s10052-012-2001-6}},
\href{http://www.arXiv.org/abs/1203.2165}{\texttt{arXiv:1203.2165}}.

\bibitem{Bern:2011ie}
Z.~Bern\hrefCMSnoop {}{ {et~al.}, ``Left-handed {$W$} bosons at the {LHC}'',}
  \textit{ Phys. Rev. D} \textbf{ 84} (2011) 034008,
  \href{http://dx.doi.org/10.1103/PhysRevD.84.034008}{\doi{10.1103/PhysRevD.84.034008}},
\href{http://www.arXiv.org/abs/1103.5445}{\texttt{arXiv:1103.5445}}.

\bibitem{Gavin:2010az}
\hrefCMSnoop {}{R.~Gavin, Y.~Li, F.~Petriello, and S.~Quackenbush, ``{FEWZ}
  2.0: A code for hadronic {Z} production at next-to-next-to-leading order'',}
  \textit{ Comput. Phys. Commun.} \textbf{ 182} (2011) 2388,
  \href{http://dx.doi.org/10.1016/j.cpc.2011.06.008}{\doi{10.1016/j.cpc.2011.06.008}},
\href{http://www.arXiv.org/abs/1011.3540}{\texttt{arXiv:1011.3540}}.

\bibitem{Alwall:2014hca}
J.~Alwall\hrefCMSnoop {}{ {et~al.}, ``{The automated computation of tree-level
  and next-to-leading order differential cross sections, and their matching to
  parton shower simulations}'',} \textit{ JHEP} \textbf{ 07} (2014) 079,
  \href{http://dx.doi.org/10.1007/JHEP07(2014)079}{\doi{10.1007/JHEP07(2014)079}},
\href{http://www.arXiv.org/abs/1405.0301}{\texttt{arXiv:1405.0301}}.

\bibitem{Chatrchyan:2008zzk}
\hrefCMSnoop {}{{CMS} Collaboration, ``The {CMS} experiment at the {CERN}
  {LHC}'',} \textit{ JINST} \textbf{ 3} (2008) S08004,
  \href{http://dx.doi.org/10.1088/1748-0221/3/08/S08004}{\doi{10.1088/1748-0221/3/08/S08004}}.

\bibitem{CMS:2009nxa}
\href {http://cdsweb.cern.ch/record/1194487}{{CMS} Collaboration,
  ``Particle--Flow Event Reconstruction in {CMS} and Performance for Jets,
  Taus, and {\MET}'',} CMS Physics Analysis Summary CMS-PAS-PFT-09-001, 2009.

\bibitem{CMS:2010byl}
\href {http://cdsweb.cern.ch/record/1247373}{{CMS} Collaboration,
  ``Commissioning of the Particle-flow Event Reconstruction with the first
  {LHC} collisions recorded in the {CMS} detector'',} CMS Physics Analysis
  Summary CMS-PAS-PFT-10-001, 2010.

\bibitem{Sjostrand:2006za}
\hrefCMSnoop {}{T.~Sj$\rm{\ddot{o}}$strand, S.~Mrenna, and P.~Z. Skands,
  ``{PYTHIA} 6.4 physics and manual'',} \textit{ JHEP} \textbf{ 05} (2006) 026,
  \href{http://dx.doi.org/10.1088/1126-6708/2006/05/026}{\doi{10.1088/1126-6708/2006/05/026}},
\href{http://www.arXiv.org/abs/hep-ph/0603175}{\texttt{arXiv:hep-ph/0603175}}.

\bibitem{z2tune}
\hrefCMSnoop {}{{CMS} Collaboration, ``Study of the underlying event at forward
  rapidity in pp collisions at $\sqrt{s}=$ 0.9, 2.76, and 7 TeV'',} \textit{
  JHEP} \textbf{ 04} (2013) 072,
  \href{http://dx.doi.org/10.1007/JHEP04(2013)072}{\doi{10.1007/JHEP04(2013)072}},
  \href{http://www.arXiv.org/abs/1302.2394}{\texttt{arXiv:1302.2394}}.

\bibitem{Alwall:2007fs}
J.~Alwall\hrefCMSnoop {}{ {et~al.}, ``Comparative study of various algorithms
  for the merging of parton showers and matrix elements in hadronic
  collisions'',} \textit{ Eur. Phys. J. C} \textbf{ 53} (2008) 473,
  \href{http://dx.doi.org/10.1140/epjc/s10052-007-0490-5}{\doi{10.1140/epjc/s10052-007-0490-5}},
\href{http://www.arXiv.org/abs/0706.2569}{\texttt{arXiv:0706.2569}}.

\bibitem{Pumplin:2002vw}
J.~Pumplin\hrefCMSnoop {}{ {et~al.}, ``{New generation of parton distributions
  with uncertainties from global QCD analysis}'',} \textit{ JHEP} \textbf{ 07}
  (2002) 012,
  \href{http://dx.doi.org/10.1088/1126-6708/2002/07/012}{\doi{10.1088/1126-6708/2002/07/012}},
  \href{http://www.arXiv.org/abs/hep-ph/0201195}{\texttt{arXiv:hep-ph/0201195}}.

\bibitem{ct10_pdf}
H.-L. Lai\hrefCMSnoop {}{ {et~al.}, ``New parton distributions for collider
  physics'',} \textit{ Phys. Rev. D} \textbf{ 82} (2010) 074024,
  \href{http://dx.doi.org/10.1103/PhysRevD.82.074024}{\doi{10.1103/PhysRevD.82.074024}},
\href{http://www.arXiv.org/abs/1007.2241}{\texttt{arXiv:1007.2241}}.

\bibitem{Alioli:2009je}
\hrefCMSnoop {}{S.~Alioli, P.~Nason, C.~Oleari, and E.~Re, ``{NLO} single-top
  production matched with shower in {POWHEG}: s- and t-channel
  contributions'',} \textit{ JHEP} \textbf{ 09} (2009) 111,
  \href{http://dx.doi.org/10.1007/JHEP02(2010)011}{\doi{10.1007/JHEP02(2010)011}},
\href{http://www.arXiv.org/abs/0907.4076}{\texttt{arXiv:0907.4076}}.

\bibitem{Re:2010bp}
\hrefCMSnoop {}{E.~Re, ``Single-top {Wt}-channel production matched with parton
  showers using the {POWHEG} method'',} \textit{ Eur. Phys. J. C} \textbf{ 71}
  (2011) 1547,
  \href{http://dx.doi.org/10.1140/epjc/s10052-011-1547-z}{\doi{10.1140/epjc/s10052-011-1547-z}},
\href{http://www.arXiv.org/abs/1009.2450}{\texttt{arXiv:1009.2450}}.

\bibitem{Czakon:2013goa}
\hrefCMSnoop {}{M.~Czakon, P.~Fiedler, and A.~Mitov, ``Total Top-Quark
  Pair-Production Cross Section at Hadron Colliders Through
  $O(\alpha^{4}_{S})$'',} \textit{ Phys. Rev. Lett.} \textbf{ 110} (2013)
  252004,
  \href{http://dx.doi.org/10.1103/PhysRevLett.110.252004}{\doi{10.1103/PhysRevLett.110.252004}},
\href{http://www.arXiv.org/abs/1303.6254}{\texttt{arXiv:1303.6254}}.

\bibitem{Kidonakis:2012db}
\hrefCMSnoop {}{N.~Kidonakis, ``{Differential and total cross sections for top
  pair and single top production}'',} in \textit{ XX Int. Workshop on
  Deep-Inelastic Scattering and Related Subjects}, p.~831.
\newblock Bonn, Germany, 2012.
\newblock \href{http://www.arXiv.org/abs/1205.3453}{\texttt{arXiv:1205.3453}}.
\newblock
\href{http://dx.doi.org/10.3204/DESY-PROC-2012-02/251}{\doi{10.3204/DESY-PROC-2012-02/251}}.

\bibitem{Agostinelli:2002hh}
\hrefCMSnoop {}{{GEANT4} Collaboration, ``{GEANT4}: A simulation toolkit'',}
  \textit{ Nucl. Instrum. Meth. A} \textbf{ 506} (2003) 250,
\href{http://dx.doi.org/10.1016/S0168-9002(03)01368-8}{\doi{10.1016/S0168-9002(03)01368-8}}.

\bibitem{Chatrchyan:2012xi}
\hrefCMSnoop {}{{CMS} Collaboration, ``Performance of {CMS} muon reconstruction
  in $\rm{pp}$ collision events at $\sqrt{s}=7$ {TeV}'',} \textit{ JINST}
  \textbf{ 7} (2012) P10002,
  \href{http://dx.doi.org/10.1088/1748-0221/7/10/P10002}{\doi{10.1088/1748-0221/7/10/P10002}},
\href{http://www.arXiv.org/abs/1206.4071}{\texttt{arXiv:1206.4071}}.

\bibitem{CMS:2013hoa}
\hrefCMSnoop {}{{CMS} Collaboration, ``{Performance of electron reconstruction
  and selection with the CMS detector in proton-proton collisions at
  $\sqrt{s}=8$\TeV}'',} \textit{ JINST} \textbf{ 10} (2015) P06005,
  \href{http://dx.doi.org/10.1088/1748-0221/10/06/P06005}{\doi{10.1088/1748-0221/10/06/P06005}},
  \href{http://www.arXiv.org/abs/1502.02701}{\texttt{arXiv:1502.02701}}.

\bibitem{rochcor}
A.~Bodek\hrefCMSnoop {}{ {et~al.}, ``Extracting muon momentum scale corrections
  for hadron collider experiments'',} \textit{ Eur. Phys. J. C} \textbf{ 72}
  (2012) 2194,
  \href{http://dx.doi.org/10.1140/epjc/s10052-012-2194-8}{\doi{10.1140/epjc/s10052-012-2194-8}},
  \href{http://www.arXiv.org/abs/1208.3710}{\texttt{arXiv:1208.3710}}.

\bibitem{Khachatryan:2010xn}
\hrefCMSnoop {}{{CMS} Collaboration, ``Measurements of inclusive $\rm{W}$ and
  $\rm{Z}$ cross sections in $\rm{pp}$ collisions at $\sqrt{s}=7$ {TeV}'',}
  \textit{ JHEP} \textbf{ 01} (2011) 080,
  \href{http://dx.doi.org/10.1007/JHEP01(2011)080}{\doi{10.1007/JHEP01(2011)080}},
\href{http://www.arXiv.org/abs/1012.2466}{\texttt{arXiv:1012.2466}}.

\bibitem{Mirkes:1994dp}
\hrefCMSnoop {}{E.~Mirkes and J.~Ohnemus, ``Angular distributions of
  {Drell-Yan} lepton pairs at the {Fermilab} {Tevatron}: Order $\alpha_S^{2}$
  corrections and Monte Carlo studies'',} \textit{ Phys. Rev. D} \textbf{ 51}
  (1995) 4891,
  \href{http://dx.doi.org/10.1103/PhysRevD.51.4891}{\doi{10.1103/PhysRevD.51.4891}},
\href{http://www.arXiv.org/abs/hep-ph/9412289}{\texttt{arXiv:hep-ph/9412289}}.

\end{thebibliography}\endgroup

\cleardoublepage \appendix\section{The CMS Collaboration \label{app:collab}}\begin{sloppypar}\hyphenpenalty=5000\widowpenalty=500\clubpenalty=5000\textbf{Yerevan Physics Institute,  Yerevan,  Armenia}\\*[0pt]
V.~Khachatryan, A.M.~Sirunyan, A.~Tumasyan
\vskip\cmsinstskip
\textbf{Institut f\"{u}r Hochenergiephysik der OeAW,  Wien,  Austria}\\*[0pt]
W.~Adam, T.~Bergauer, M.~Dragicevic, J.~Er\"{o}, M.~Friedl, R.~Fr\"{u}hwirth\cmsAuthorMark{1}, V.M.~Ghete, C.~Hartl, N.~H\"{o}rmann, J.~Hrubec, M.~Jeitler\cmsAuthorMark{1}, W.~Kiesenhofer, V.~Kn\"{u}nz, M.~Krammer\cmsAuthorMark{1}, I.~Kr\"{a}tschmer, D.~Liko, I.~Mikulec, D.~Rabady\cmsAuthorMark{2}, B.~Rahbaran, H.~Rohringer, J.~Schieck, R.~Sch\"{o}fbeck, J.~Strauss, W.~Treberer-Treberspurg, W.~Waltenberger, C.-E.~Wulz\cmsAuthorMark{1}
% ...
\vskip\cmsinstskip
% ...
1:~~Also at Vienna University of Technology, Vienna, Austria\\
2:~~Also at CERN, European Organization for Nuclear Research, Geneva, Switzerland\\
% ...

\end{sloppypar}
\end{document}